\documentclass{aa}  

\usepackage{graphicx}
\usepackage{txfonts}
\usepackage{natbib}
\usepackage{amsmath} %for dealing with mathematics,
\usepackage{multirow}
\usepackage{booktabs}
\usepackage{float}
\usepackage{hyperref} %%% for linking the cross references
\usepackage{graphics} %%%for dealing with figures.
\usepackage{subfig} %%for getting the subfigures e.g., "Figure 1a and 1b" etc.
\usepackage{xcolor}
\usepackage{enumitem}
\usepackage{amsmath}
\usepackage{bm}
\usepackage{mathtools}
\usepackage{soul}

\begin{document}

   \title{Planet Earth in reflected and polarized light}

   \subtitle{II. Refining contrast estimates for rocky exoplanets with ELT and HWO}

   \author{Giulia Roccetti
          \inst{1,2}
          \and
          Michael F. Sterzik
          \inst{1}
          \and
          Julia V. Seidel
          \inst{3,4}
          \and
          Claudia Emde
          \inst{2,5}
          \and
          Mihail Manev
          \inst{2}
          \and
          Stefano Bagnulo
          \inst{6}
          }

   \institute{
    European Southern Observatory, Karl-Schwarzschild-Straße 2, 85748, Garching near Munich, Germany\\
    \email{giulia.roccetti@eso.org}
    \and
    Meteorologisches Institut, Ludwig-Maximilians-Universität München, Munich, Germany
    \and
    European Southern Observatory, Santiago, Chile
    \and
    Laboratoire Lagrange, Observatoire de la Côte d’Azur, CNRS, Universit\'e Côte d’Azur, Nice, France
    \and
    Rayference, Rue d’Alost 7, 1000, Bruxelles, Belgium
    \and
    Armagh Observatory and Planetarium, College Hill, Armagh BT61 9DG, Northern Ireland, UK}
    
   \date{Received xx; accepted xx}

% \abstract{}{}{}{}{} 
% 5 {} token are mandatory
 
  \abstract{The characterization of nearby rocky exoplanets will become feasible with the next generation of telescopes, such as the Extremely Large Telescope (ELT) and the mission concept Habitable Worlds Observatory (HWO). Using an improved model setup, we aim to refine the estimates of reflected and polarized light contrast for a selected sample of rocky exoplanets in the habitable zones of nearby stars. We perform advanced 3D radiative transfer simulations for Earth-like planets orbiting G-type and M-type stars. Our simulations incorporate realistic, wavelength-dependent surface albedo maps and a detailed cloud treatment, including 3D cloud structures and inhomogeneities, to better capture their radiative response. These improvements are based on Earth observations. We present models of increasing complexity, ranging from simple homogeneous representations to a detailed Earth-as-an-exoplanet model. Our results show that averaging homogeneous models fails to capture Earth's full complexity, especially in polarization. Moreover, simplistic cloud models distort the representation of absorption lines at high spectral resolutions, particularly in water bands, potentially biasing atmospheric chemical abundance estimates. Additionally, we provide updated contrast estimates for observing rocky exoplanets around nearby stars with upcoming instruments such as ANDES and PCS at the ELT. Compared to previous studies, our results indicate that reflected light contrast estimates are overestimated by a factor of two when simplified cloud and surface models are used. Instead, measuring the fractional polarization in the continuum and in high-contrast, high-resolution spectra may be more effective for characterizing nearby Earth-like exoplanets. These refined estimates are essential for guiding the design of future ELT instruments and the HWO mission concept.
  }


   \maketitle
%
%-------------------------------------------------------------------

\section{Introduction}
\label{sec:intro}

The search for Earth-like exoplanets and the quest to characterize their atmospheres remain among the most compelling goals of modern astrophysics. However, a major challenge in directly imaging and characterizing rocky exoplanets is the extreme contrast between the planet and its much brighter host star. Overcoming this challenge requires innovative observational techniques and a deeper understanding of planetary light scattering.\\
\noindent Currently, most mature exoplanet atmosphere characterization relies on transmission spectroscopy, which is limited to transiting planets. Although highly successful for studying the atmospheric dynamics of hot Jupiters during transit (e.g., \citealt{Nortmann2025, Seidel2025}) and secondary eclipse (e.g., \citealt{Pino2020, CostaSilva2024}) and their atmospheric chemical composition (e.g., \citealt{Snellen2008, Prinoth2024b}), this method faces significant challenges when applied to rocky exoplanets. Temperate rocky exoplanets are less likely to transit due to their typically longer orbital periods, making transmission spectroscopy impractical for a broad population of potentially habitable worlds. Furthermore, their atmospheres are generally thinner than those of Jupiter-like planets, making atmospheric characterization even more challenging. In addition,  in-transit radial velocity changes for longer period planets (like warm Jupiters) are too small to directly separate the planetary signal's wavelength shift from stellar and telluric lines, further complicating the detection of atmospheric absorption lines \citep{borsa2019, seidel2020b, seidel2020c, Prinoth2024a}. Moreover, transmission spectroscopy primarily probes the upper atmospheric layers, providing limited information about surface conditions and lower atmospheric composition, both crucial for assessing habitability. Importantly, in this context, habitability refers specifically to the potential for surface liquid water, not necessarily to conditions sufficient for supporting life.\\
\noindent Reflected light observations offer a promising alternative, enabling the study of both transiting and non-transiting exoplanets. By analyzing the light scattered by the atmosphere and reflected off the planet’s surface, it is possible to retrieve key properties such as albedo across different wavelengths and phase angles \citep{Roccetti2025a}, which can reveal the presence of clouds, oceans, and ice \citep{turbet2016}. Additionally, spectral features in reflected light provide direct constraints on atmospheric composition, allowing for the detection of key molecules such as O$_2$, H$_2$O, CH$_4$, CO$_2$, and potentially CO. These observations can offer a more comprehensive view of an exoplanet’s climate, surface conditions, and habitability.\\
\noindent The upcoming RISTRETTO \citep{Lovis2022} spectrograph at the Very Large Telescope (VLT) is designed for detecting and analyzing exoplanetary atmospheres in reflected light, with a primary focus on the temperate rocky planet Proxima b \citep{Bugatti2024}. This scientific mission leverages the synergy between a high-contrast adaptive optics (AO) system and high-resolution spectroscopy and will serve as a precursor to the Extremely Large Telescope (ELT). The ELT, with its unprecedented light-collecting capacity and angular resolution, will pioneer the detection of reflected light from rocky exoplanets. Although current attempts have mostly yielded upper limits \citep{charbonneau1999, collier-cameron1999, collier-cameron2002, rodler2013, martins2015, hoeijmakers2018, scandariato2021, spring2022}, high-contrast, high-resolution (HCHR) observations with the ELT's ANDES instrument are expected to achieve contrasts on the order of 10$^{-7}$ within a few tens of nights \citep{palle2023}. A promising golden sample for future observations includes Proxima b, GJ 682 b, Wolf 1061 c, GJ 273 b and Ross 128 b, all orbiting M dwarfs. Observationally, M dwarfs are favorable targets because their close-in habitable zones \citep{kasting1993, Selsis2007, Kopparapu2013} enable the detection and characterization of more transiting planets on shorter orbits. Additionally, other ELT instruments like HARMONI may also allow the characterization of the atmosphere of Proxima b \citep{vaughan2024}, while the proposed PCS instrument \citep{kasper2021} will leverage its extreme AO system and spectrograph for imaging rocky exoplanets. \\
\noindent Rocky planets orbiting M dwarfs may face significant challenges in retaining substantial atmospheres due to intense stellar activity and irradiation \citep{luger&barnes2015, dong2018}. Recent JWST results offer mixed findings: 55 Cancri e is proposed to have an atmosphere \citep{hu2024}, while TRAPPIST-1 b and c showed upper limits on the absence of thick atmospheres \citep{greene2023, zieba2023}. Detecting atmospheres around habitable zone planets remains difficult due to stellar contamination, as shown in the case of LHS 1140 b \citep{cadieux2024}. \\
\noindent G-type stars, with their relatively stable stellar activity, are more promising targets for finding true Earth analogs. Ongoing efforts like the Terra Hunting Experiment (THE, \citealt{hall2018}) are focused on detecting Earth-mass planets around G-type stars through precise radial velocity measurements. Moreover, the PLATO mission \citep{Rauer2024} will play a key role in detecting and characterizing Earth-sized planets around Sun-like stars by leveraging high-precision photometry to measure their transits and constrain their bulk properties. Additionally, the ARIEL space mission \citep{Tinetti2018} will begin characterizing a subset of temperate sub-Neptunes and super-Earths \citep{Edwards&Tinetti2022}, providing valuable input for the target selection of upcoming space missions like the mission concepts Habitable World Observatory (HWO, \citealt{US_decadal_survey}) and Large Interferometer For Exoplanets (LIFE, \citealt{Quanz2022}), which aim to search for and characterize rocky exoplanets. HWO, for instance, will aim to image and study the reflected light of exoplanets, while LIFE will focus on thermal emission to provide insights into atmospheric pressure-temperature profiles and molecular signatures \citep{alei2024}. \\
\noindent A complementary approach to characterize distant worlds through spectroscopy and direct imaging is by measuring their degree of polarization in reflected light. The light reflected by an exoplanet becomes partially linearly polarized due to atmospheric scattering and surface reflection \citep{stam2008}. Previous modelling efforts by \cite{stam2008, karalidi2012, munoz2015, emde2017, trees2019, trees2022, gordon2023, vaughan2023} have demonstrated how polarization can aid in characterizing Earth-like exoplanets and distinguishing between different atmospheric and surface properties, which intensity-only simulations cannot achieve. \cite{Roccetti2025a} performed detailed simulations of reflectance and polarized spectra, as well as phase curves, for Ocean and Earth-like planetary scenarios. Their results suggest that the polarization fraction for an Earth-like planet ranges between 10\% and 30\%, depending on the wavelength and phase angle. Studying polarized light offers a valuable way to enhance the contrast between the planet and its host star, as most F-, G-, and K-type stars emit nearly unpolarized light \citep{cotton2017}. However, this method comes with the trade-off of reduced sensitivity. Furthermore, polarization measurements help break degeneracies in atmospheric retrievals by distinguishing between clouds and surface features across different models \citep{karalidi2012}. Unlike intensity-based observations, polarized light is not affected by transmission through Earth’s atmosphere, as polarization arises from scattering processes with molecules or surface reflections, eliminating the need for telluric correction.\\
\noindent \cite{Roccetti2025a} introduced an improved cloud and surface modeling framework for simulating the reflected and polarized light of rocky exoplanets, highlighting the impact of realistic cloud and surface modeling. Their study demonstrated that neglecting sub-grid cloud variability leads to a significant overestimation of the planet’s overall reflectance. Additionally, the vegetation red edge (VRE) feature is largely overestimated if surface albedo maps do not account for wavelength-dependent variations of complex mixture of different surface materials.\\
\noindent Building on this refined modeling approach, this second paper in the series extends contrast estimations for the reflected and polarized light of future rocky exoplanet observations. Specifically, we construct homogeneous planetary models and evaluate how their spectral and phase curve features compare to the more detailed, realistic models presented in \cite{Roccetti2025a}. Using the defined golden sample of rocky exoplanets expected to be observed in reflected light by ANDES \citep{palle2023}, we provide improved contrast estimates and compare them with values reported in the literature. This analysis assesses the impact of detailed cloud and surface modeling in 3D radiative transfer simulations. Additionally, we explore contrast predictions for polarized observations and evaluate their feasibility with the upcoming ELT and HWO. These results are crucial for guiding the design of future instruments dedicated to the characterization of rocky exoplanets.

%--------------------------------------------------------------------
\section{Reflected and polarized light contrasts}
\label{sec:contrasts}

\cite{palle2023} identified a golden sample of non-transiting rocky exoplanets orbiting nearby M-type stars, for which ANDES is expected to detect reflected light within a few tens of nights. This sample consists of the five most promising targets in terms of reflected light signal-to-noise ratio (SNR): Proxima b, GJ 273 b, Wolf 1061 c, GJ 682 b, and Ross 128 b. With the planned launch of PLATO in 2026, serving as a key pathfinder survey, the target list is expected to expand in the coming years. Building on the golden sample, we include the recently discovered Barnard b \citep{Gonazalez-Hernandez2024}, a sub-Earth-mass exoplanet orbiting the closest single star to the Sun. Since all six of these rocky exoplanets orbit M-dwarfs, we extend our study to include a comparison with a G-type star by estimating the contrast of potential Earth-like exoplanet in the habitable zone of Alpha Centauri A. Notably, \cite{Wagner2021} reported the detection of a point-like source in this system, which could be attributed to an exoplanet, exozodiacal dust, or an instrumental artifact, pushing the current exoplanet imaging mass detection limits. \\
\noindent Our goal is to provide refined flux contrast estimates for these nearby planetary systems by improving the modeling of clouds and surface albedo in reflected light calculations. In general, the contrast between an exoplanet and its host star in reflected light is
\begin{equation}
    C_{\text{flux}} = \frac{F_\text{p}}{F_{\star}} = \left(\frac{R_\text{p}}{d \sin{\theta_\text{sep}}}\right)^2 A_\text{g} \cdot g(\alpha)
\end{equation}
where $F_\text{p}$ and $F_{\star}$ are the fluxes of the planet and the star, respectively, $R_\text{p}$ is the radius of the planet, $d$ is the distance from the Earth, $A_\text{g}$ is the geometric albedo of the planet, $g (\alpha$) is the phase function and $\alpha$ the phase angle (e.g., the angle between the direction to the star and the direction to the observer as seen from the planet). The angular separation $\theta_{\text{sep}}$ of the star-planet system is dependent on the phase angle, and we report values for $\alpha$ = 90\degr, corresponding to the maximum elongation of the planet, to allow for a direct comparison with the contrast estimates presented in \cite{palle2023}. However, ANDES can operate at smaller inner working angles and observe planets at phase angles below 90\degr, which are more favorable for detecting reflected light, thereby enhancing contrast \citep{palle2023}. \\
\noindent In \cite{Roccetti2025a}, we present extensive sensitivity studies on estimating planetary reflectance, defined as the product of the geometric albedo and the phase function: $R = A_\text{g} \cdot g(\alpha)$. Since reflectance is independent of the stellar spectrum, the flux contrast can be determined using the planetary radius and angular separation for different planetary systems. In Table \ref{tab:scale_factor}, we provide typical values of the scale factor ($s$), defined as
\begin{equation}
    s = \left(\frac{R_\text{p}}{d\sin{\theta_{\text{sep}}}}\right)^2 .
    \label{eq:scale_factor}
\end{equation}
\begin{table*}[h]
    \caption{\label{tab:scale_factor} Scale factor for exoplanets orbiting nearby M-type stars and a potential Earth-like planet around Alpha Cen A.}
    \centering
    \resizebox{0.79\textwidth}{!}{%%
    \begin{tabular}{ccccccc}
    \hline
    Name & SpecType (T$_{\rm eff}$) & d [pc] & $\theta_{\text{sep}} [\text{mas}]$ & V [mag] & $R_{\text{p}} [R_{\oplus}]$ & s\\
    \hline
    Proxima Cen b & M (2900 K) & 1.30 & 37.3 & 11.01 & 1.07 & 8.85 $\cdot 10^{-7}$\\
    Ross 128 b    & M (3163 K) & 3.37 & 14.7 & 11.12 & 1.15 & 9.79 $\cdot 10^{-7}$\\
    GJ 273 b      & M (3382 K) & 3.80 & 24.0 & 9.84  & 1.64 & 5.88$\cdot 10^{-7}$\\
    Wolf 1061 c   & M (3309 K) & 4.31 & 20.7 & 10.10 & 1.81 & 7.48$\cdot 10^{-7}$\\
    GJ 682 c      & M (3237 K) & 5.01 & 16.0 & 10.94 & 2.11 & 1.26$\cdot 10^{-6}$\\
    Barnard b     & M (3195 K) & 1.83 & 12.9  & 9.51  & 0.76 & 1.88 $\cdot 10^{-6}$\\
    Alpha Cen A   & G (5804 K) & 1.34 & 747  & 0.01  & 1.0  & 1.82$\cdot 10^{-9}$\\
    \hline
    \end{tabular}
    }
\end{table*}
\noindent In addition to the reflected light contrast, we also want to introduce the contrast in polarization. The incident starlight reaching the planet is expected to be nearly unpolarized, with the disk-integrated sunlight exhibiting a polarization level of approximately $10^{-6}$ \citep{kemp1987}. Conversely, light reflected from a planet’s surface or scattered within its atmosphere can be polarized at levels of several tens of percent. Rayleigh scattering by atmospheric molecules polarizes light, though multiple scattering with clouds and aerosols can depolarize previously polarized photons. Similarly, the ocean glint produces strong linear polarization, while other surface types may depolarize light. Polarized light observations enhance the contrast between the planet and its host star and offer advantages over total flux measurements. Unlike absolute intensity, polarization is a relative measurement, independent of the star’s type or distance. \\
\noindent In polarization, the reflected light contrast can be expressed as
\begin{equation}
    C_{\text{pol}} = C_{\text{flux}} \cdot P,
\end{equation}
where $P$ is the degree of linear polarization normalized between 0 and 1. The polarization contrast is function of $\alpha$ \citep{Buenzli&Schmid2009A}.\\
\noindent While it is true that the polarized contrast is lower than the contrast in intensity alone, polarimetric differential imaging (DPI) can greatly push the sensitivity down to few order of magnitudes due to fast modulation. The Zurich IMaging POLarimeter (ZIMPOL), the visible focal plane instrument of SPHERE can, in principle, reduce the achievable contrast from 10$^{-4}$ (with Adaptive Optics alone) to 10$^{-8}$ in polarization \citep{hunziker2020}. This advantage is particularly significant for planets with smaller angular separations from their host star, as polarimetric observations can reduce, or, in principle, cancel, speckle noise around the coronagraph. \cite{beuzit2019} conducted polarimetric observations of a sample of targets and demonstrated that ZIMPOL achieves polarization contrast detection limits much deeper than those of intensity-based observations. For Alpha Cen A, polarization lowered the achievable contrast from 10$^{-5}$ to 10$^{-7}$ close to the star, at an angular separation of 0.35~arcsec, and from 10$^{-7}$ to 10$^{-8}$ at the wider separation angle of 1.5~arcsec.

\section{3D radiative transfer simulations}
\label{3D_rad_tran}

We perform 3D radiative transfer simulations using MYSTIC \citep{mayer2009}, the Monte Carlo code for the phYsically correct Tracing of photons in Cloudy atmospheres, which is part of the libRadtran library \citep{mayer2005, emde2016}. MYSTIC incorporates the Absorption Lines Importance Sampling (ALIS) method \citep{emde2011}, enabling fast calculations of high-resolution spectra by tracing photons at a single wavelength. Moreover, the variance reduction method VROOM \citep{buas2011} is used to correctly simulate clouds. \cite{emde2017} adapted MYSTIC to simulate disk-integrated properties of the Earth as an exoplanet in polarization, accounting for surface reflection, multiple scattering by molecules, aerosol particles, cloud droplets, and ice crystals.\\
\noindent Building on this foundation, \cite{Roccetti2025a} further advanced the modeling framework by introducing the capability to simulate fully inhomogeneous and realistic planets. This includes a new treatment of cloud sub-grid variability and inhomogeneities through the 3D Cloud Generator (3D CG) algorithm\footnote{\url{https://github.com/giulia-roccetti/3D_Cloud_Generator}}. The 3D CG employs 3D cloud fields from the ERA5 reanalysis dataset, which provides atmospheric data on a global grid of 1440 × 721 horizontal pixels and 37 vertical levels. For each grid cell, ERA5 provides liquid water content, ice water content, and cloud cover. Although the ERA5's spatial resolution ($\sim$31 km or 0.25\degr) is more than adequate for exoplanet modeling, \cite{Roccetti2025a} demonstrated that introducing sub-pixel cloud variability significantly affects disk-integrated reflectance and polarization spectra and phase curves. To address this, the 3D CG redistributes the liquid and ice water content within each ERA5 grid cell into sub-grid structures, generating patchier cloud distributions. This allows more photons to reach the surface and mitigates the over-smoothing of cloud effects at coarse spatial resolution. The algorithm conserves both the in-cloud optical thickness and total planetary cloud cover. A specified vertical overlap scheme is applied to distribute sub-grid clouds vertically. \cite{Roccetti2025a} found that the algorithm converges when each ERA5 grid cell is divided into nine sub-pixels (a zoom-in factor of 3), with no significant differences observed between maximum-random and exponential-random vertical overlap schemes. Therefore, all 3D CG simulations in this work are performed using a x3 zoom with exponential-random overlap.\\
\noindent The improved modeling framework presented in \cite{Roccetti2025a} also includes the implementation of wavelength-dependent surface albedo maps using HAMSTER \citep{Roccetti2024}. Additionally, the framework incorporates more sophisticated surface treatments, allowing Lambertian surfaces with spectral albedo variations and oceans modeled with bidirectional reflectance distribution functions (BRDFs) or bidirectional polarization distribution functions (BPDFs), to allow a correct treatment of the ocean glint.\\
\noindent The results and sensitivity studies presented in \cite{Roccetti2025a} serve as a starting point for this work. Here, we assess whether a linear combination of homogeneous planet models can accurately reproduce the ground-truth Ocean and Earth-like planet scenarios explored in \cite{Roccetti2025a}. Furthermore, we investigate the impact of advanced 3D inhomogeneous radiative transfer simulations on estimating the contrast in reflected and polarized light for the golden sample of rocky exoplanets orbiting nearby stars \citep{palle2023}.

\subsection{Homogeneous planets model setup}
\label{sec:hom_planets}

We construct homogeneous, cloud-free planetary models by incorporating wavelength-dependent surface albedo properties characteristic of four distinct surface types: desert, forest, polar ice cap, and ocean, all beneath an Earth-like atmosphere with US standard atmospheric properties \citep{anderson1986}. The wavelength-dependent surface albedo properties are extracted from HAMSTER \citep{Roccetti2024}. Specifically, we use the typical reflectance spectrum of the Amazon rainforest region in HAMSTER for the forest planet. For the desert planet, we adopt the wavelength-dependent surface albedo from the Australian desert dataset. For the polar region, we use the boreal summer Antarctica spectrum as a benchmark.\\
\noindent For the ocean surface, while it lacks strong wavelength-dependent features, we account for ocean glint reflection by implementing the BRDF in reflected light and the BPDF in polarized light, assuming a constant surface wind speed of 10\,m\,s$^{-1}$. To approximate an Earth-like planet scenario, we construct a linear combination of these homogeneous planets, assuming the Earth's surface composition consists of 70\% ocean, 10\% forest, 10\% desert, and 10\% polar regions.\\
\noindent To simulate homogeneous cloudy planets, we retain the same wavelength-dependent surface properties while introducing an idealized homogeneous cloud field based on the properties detailed in \cite[Tables 1 and 2]{Roccetti2025a}. Specifically, we assume a 46\% cloud cover, with a liquid water cloud optical depth of 6.51 at a bottom altitude of 1.59~km. To ensure consistency across different horizontal resolutions, we first generate a cloud field matching the 3D CG resolution with a zoom-in factor of ×3 \citep{Roccetti2025a}, resulting in a grid box size of approximately 9~km. The cloud layer is set to a 1~km vertical extent, and we calculate the in-cloud liquid water content (LWC) to maintain the prescribed liquid water cloud optical depth. The cloud effective droplet radius is also fixed at 8.99 $\mu$m, from the averaged properties of the ERA5 reanalysis product \citep{Hersbach2020} found in \cite{Roccetti2025a}. In these homogeneous planet models, 60\% of the sub-grid cells are randomly assigned to be cloudy, resulting in horizontally patchy cloud structures. Also for the Earth-like cloudy scenario, we build it as a linear combination of the cloudy homogeneous simulations, maintaining the assumed surface composition of 70\% ocean, 10\% forest, 10\% desert, and 10\% polar regions.

\subsection{Setups for models of increasing complexity}
\label{sec:models_definition}

By building models of increasing complexities, we assess the impact of the improved cloud and surface modelling approaches presented in \cite{Roccetti2025a} on the simulated spectra and phase curves compared to homogeneous planet simulations. To perform this comparison, we use the same 3D radiative transfer code, MYSTIC, and the same grid size. We simulate models of different complexities, from uniform surface and clouds to more complex and inhomogeneous cases. The increasing complexity scales as follows:
\begin{itemize}
    \item uniform surface (constant surface albedo of 0.3, not wavelength dependent) and uniform liquid water clouds (fully cloudy layer with $\tau$ = 6.51, $r_{\rm eff}$ = 8.99~$\mu$m and altitude thickness 1\,km from 1.59 to 2.59\,km);
    \item uniform surface (as above) and a homogeneous liquid water cloud layer with 46\% patchy cloud cover (making the previous cloud layer patchy and redistributing the LWC among only cloudy pixels);
    \item linear combination of surfaces, taking a representative spectra of a forest, a desert, a polar region and the ocean (including BRDF and BPDF), and averaging them as 70\% ocean, 10\% forests, 10\% deserts and 10\% polar to reproduce the Earth, and patchy liquid water clouds with 46\% cloud cover (as above);
    \item linear combination of surfaces (as above) with two cloud layers, the patchy liquid water clouds and the patchy ice water clouds. For the ice water clouds we use 54\% of cloud cover, $\tau$ = 0.63, altitude range from 4.34 to 5.34\,km and $r_{\rm eff}$ = 46.9~$\mu$m taken from \citep[Tables 1 and 2]{Roccetti2025a};
    \item ocean surface (including BRDF and BPDF) with the 3D CG clouds, including their liquid water and ice water clouds and their 1$\sigma$ spread, this scenario obviously includes the effects of an ocean glint;
    \item Earth-like scenario including ocean surface treated with the BRDF and BPDF (but ocean glint almost always covered by land) and hyperspectral albedo maps with the 3D CG clouds including their liquid water and ice water clouds and their 1$\sigma$ spread.
\end{itemize}

\subsection{Setup for high spectral resolution simulations}

With the same model setup as in \cite{Roccetti2025a}, we run high spectral resolution simulations at ANDES resulution R = 100\,000 to study the effect of different surface and cloud properties not only on the continuum, but also on the absorption lines. Upcoming instruments at the ELT, such as ANDES, will allow to image the closer rocky exoplanets orbiting M dwarfs using HCHR observations. We study in detail the O$_2$-A band around 780~nm and the H$_2$O absorption lines in the Y band, between 920 and 950~nm. To perform high spectral resolution simulations, we couple the Atmospheric Radiative Transfer Simulator (ARTS version 2.2; \citealt{buehler2005}, \citealt{eriksson2011}) with MYSTIC. ARTS provides accurate line-by-line absorption calculations for molecular species, ensuring precise spectral resolution across a wide wavelength range. The computed absorption coefficients are then used as input for MYSTIC, which simulates the 3D radiative transfer. 

\subsection{M-dwarf simulations model setup}

We extend the comparison between the Ocean and Earth-like planet scenarios from \cite{Roccetti2025a} to an exoplanet orbiting an M dwarf star. Using the same models, we simulate reflected and polarized light spectra across a wavelength range of 400–2500~nm at a spectral resolution of 1~nm using the REPTRAN absorption parametrization \citep{gasteiger2014}. While the stellar spectrum is updated as an input, the resulting spectra are only affected by the planet's surface, atmosphere, and cloud properties, and not by the stellar spectrum. As expected, the reflected light contrast decreases toward the near-infrared (NIR), but this analysis provides valuable predictions for upcoming NIR observations. Moreover, the NIR contains a higher density of absorption features compared to the visible range, enabling the detection of key atmospheric species such as H$_2$O, CO$_2$, CH$_4$, and O$_2$, which are crucial for exoplanet characterization.

\section{From homogeneous to realistic Earth-like planets}
\label{sec:results}

\subsection{Homogeneous planets spectra and phase curves}
\label{subsec:hom_planets}

\begin{figure*}
    \centering
    \includegraphics[width=1\linewidth]{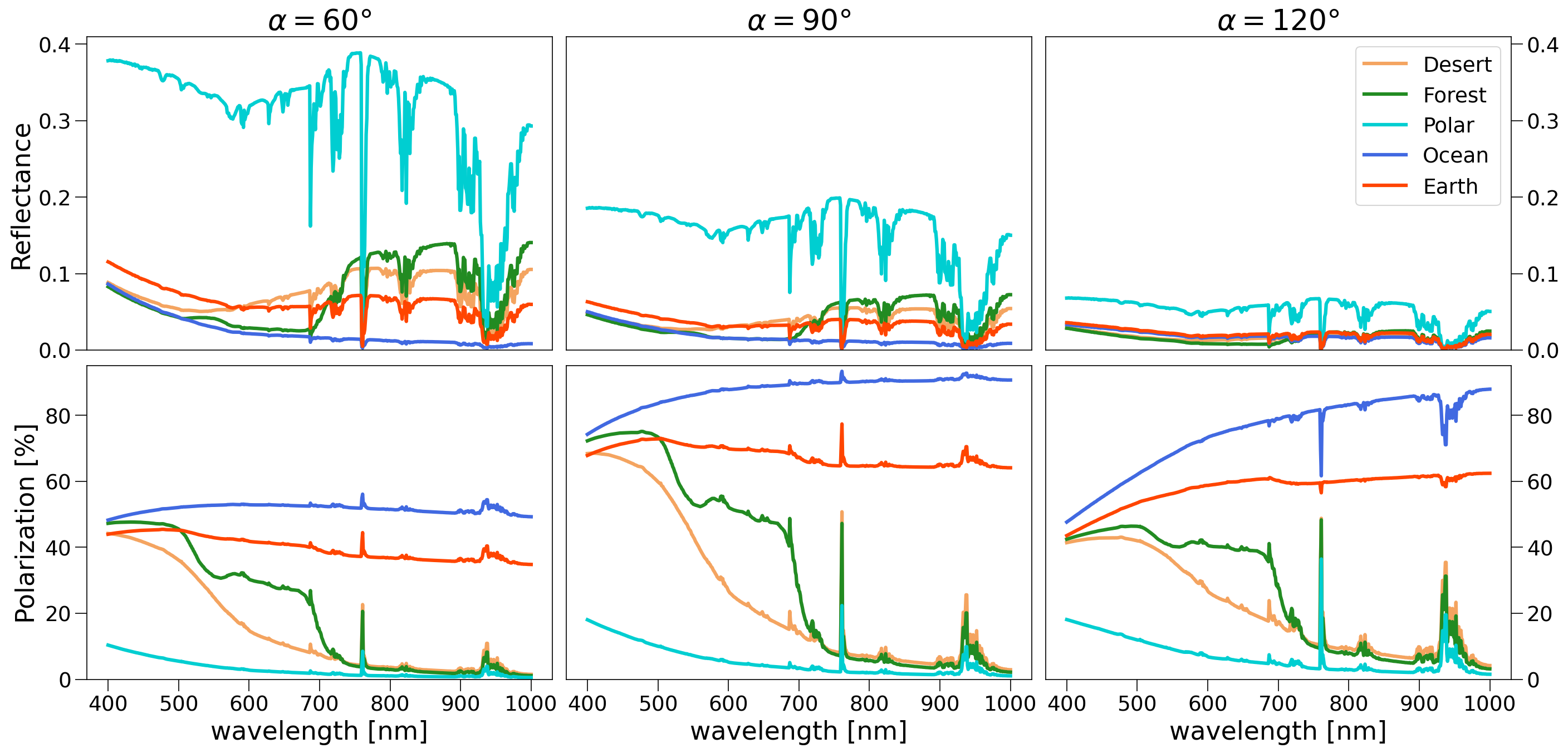}
    \caption{Reflected light (top row) and polarized light (bottom row) spectra for various homogeneous, cloud-free planets with different surface types. The wavelength-dependent spectral features of desert, forest, and polar surfaces are taken from HAMSTER \citep{Roccetti2024} and modeled as Lambertian reflectors, while the ocean surface incorporates BRDF and BPDF treatments. Each column corresponds to spectra at different phase angles $\alpha$: 60, 90, 120\degr.}
    \label{fig:spectra_hom}
\end{figure*}

\begin{figure*}
    \centering
    \includegraphics[width=1\linewidth]{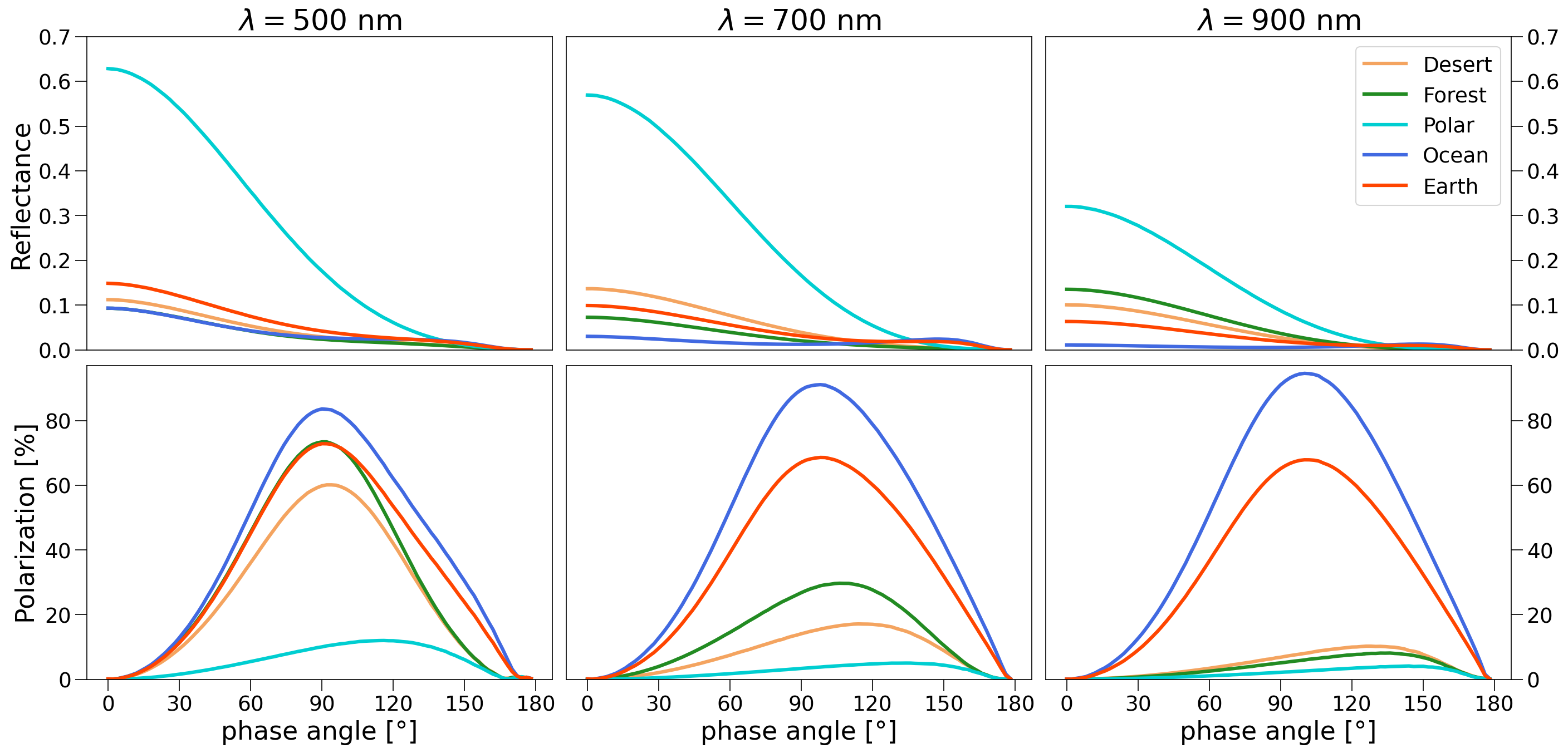}
    \caption{Reflected light (top row) and polarized light (bottom row) phase curves showing homogeneous cloud-free planets. The wavelength-dependent spectral features of desert, forest, and polar surfaces are taken from HAMSTER \citep{Roccetti2024} and modeled as Lambertian reflectors, while the ocean surface incorporates BRDF and BPDF treatments. Different columns refer to different wavelengths ($\lambda$): 500, 700, 900~nm.}
    \label{fig:phase_curves_hom}
\end{figure*}

We perform simulations for homogeneous planets using wavelength-dependent albedo properties from HAMSTER \citep{Roccetti2024}. As demonstrated in \cite{Roccetti2025a}, accurately modeling surface reflectance, whether for forests, deserts, or other surface types, significantly impacts the planet’s total reflectance, particularly in the VRE region. \cite{Roccetti2025a} showed that previous models substantially overestimated the VRE because they represented vegetated surfaces using the laboratory-measured reflectance of a single leaf. However, a forest is a far more complex environment, with its spectral signature arising from a combination of leaves, soil, and other materials. By incorporating HAMSTER into our radiative transfer simulations, \citep{Roccetti2025a} demonstrates that the increase in reflectance around 750~nm due to the VRE is notably smaller than previously estimated. This finding helps explain why the observed intensity of the VRE in Earth as an exoplanet intensity observations is weaker than previously expected, as seen in Earthshine studies \citep{Montanes-Rodriguez2006}.\\
\noindent Here, we present reference spectra and phase curves for homogeneous, cloud-free planets with different surface types in both reflected and polarized light. The ocean planet includes BRDF and BPDF treatments, while other land surface types are derived from the HAMSTER hyperspectral albedo maps dataset \citep{Roccetti2024} and treated as Lambertian surfaces. The Earth-like case, shown in orange, is modeled as a linear combination of ocean, desert, forest, and polar surface types.\\
\noindent In the absence of clouds, distinct surface-dependent features emerge both in the spectra (Fig. \ref{fig:spectra_hom}) and phase curves (Fig. \ref{fig:phase_curves_hom}). The polar ice cap planet (cyan model) exhibits high reflectance across all wavelengths, while the forest model (green) shows a pronounced VRE feature, which is clearly visible in the spectral reflectance of a purely forested planet. Polarization spectra, however, show the opposite trend. The ocean planet (dark blue) exhibits strong polarization due to the ocean glint effect, while the VRE feature manifests as a steep decline in polarization between 700 and 800~nm. Additionally, phase curves in polarization reveal a shift in the peak of the polarization curve depending on surface properties. For an ocean planet, the polarization peak occurs at almost 90\degr only at 500~nm, but at higher phase angles at 700 and 900~nm. The polarization peak for an ocean surface without atmosphere would be at 106\degr (double the Brewster angle for water). For the ocean planet with a Rayleigh atmosphere it is a mixture of effects: at shorter wavelengths Rayleigh scattering dominates and the peak is at 90\degr, at longer wavelengths surface interactions are decisive and the peak gets closer to 106\degr. For a forest planet, the peak shifts to larger phase angles compared to an ocean planet, with a wavelength-dependent trend. At $\lambda$ = 500~nm, where forests are darker and Rayleigh scattering dominates, the polarization peak remains close to $\alpha$ = 90\degr. However, at longer wavelengths ($\lambda$ = 700 and 900~nm), the increased reflectance of forests after the VRE causes the polarization peak to shift to approximately $\alpha$ = 110 and 130\degr, respectively.

\subsection{Patchy clouds over homogeneous surfaces}

\begin{figure*}
    \centering
    \includegraphics[width=1\linewidth]{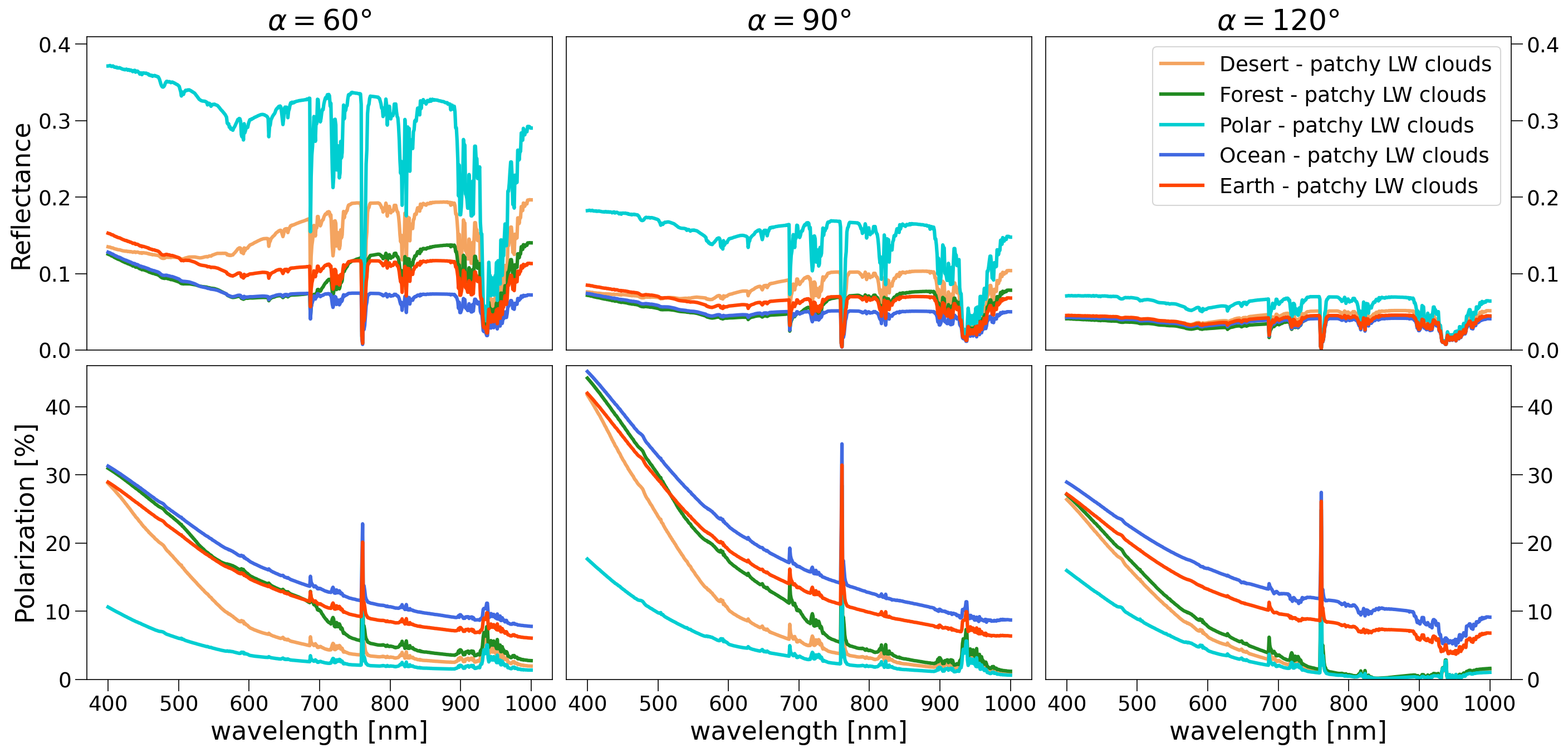}
    \caption{Reflected light (top row) and polarized light (bottom row) spectra for various homogeneous planets with different surface types and homogeneous clouds. The wavelength-dependent spectral features of desert, forest, and polar surfaces are taken from HAMSTER \citep{Roccetti2024} and modeled as Lambertian reflectors, while the ocean surface incorporates BRDF and BPDF treatments. Each column corresponds to spectra at different phase angles $\alpha$: 60, 90, 120\degr.}
    \label{fig:spectra_hom_clouds}
\end{figure*}

\begin{figure*}
    \centering
    \includegraphics[width=1\linewidth]{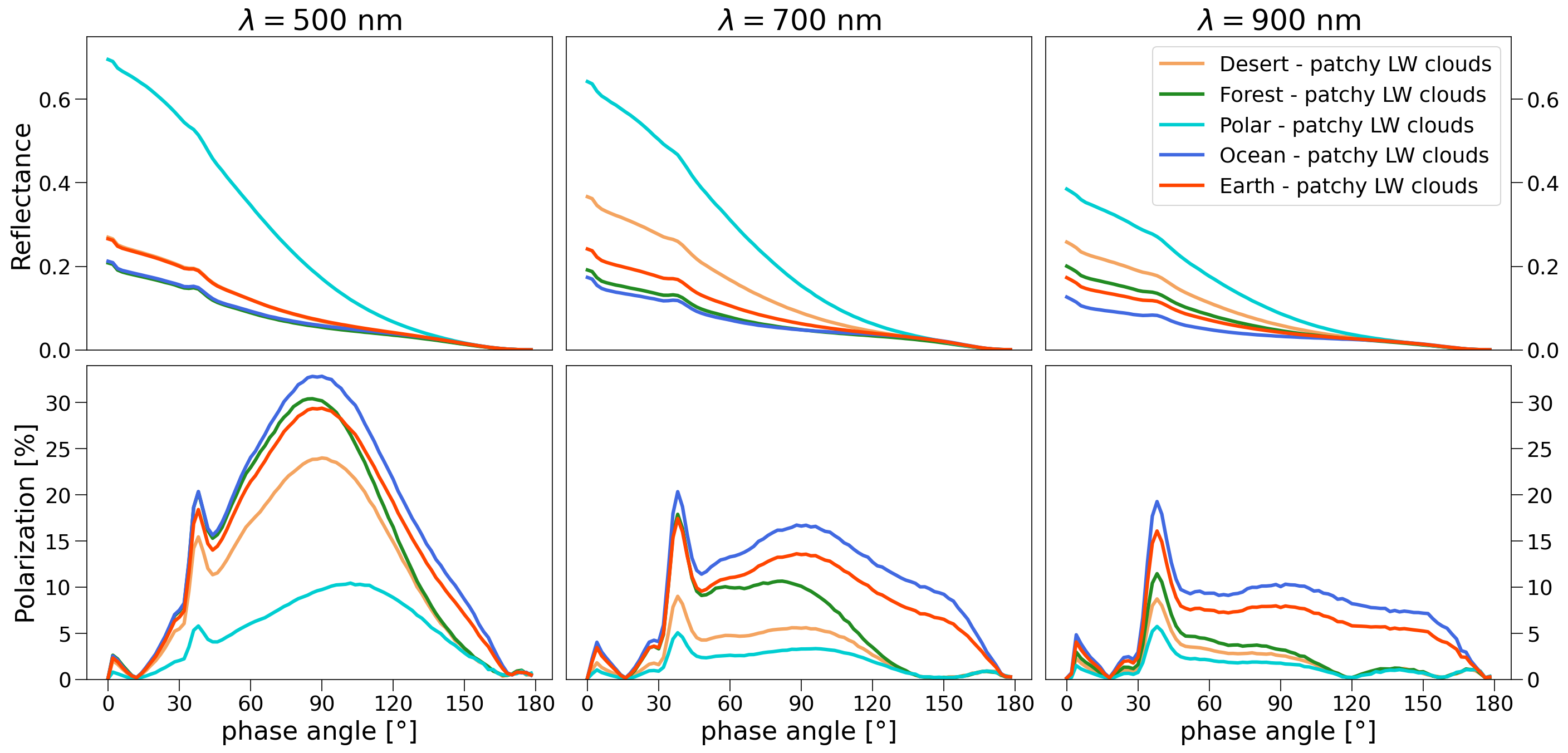}
    \caption{Reflected light (top row) and polarized light (bottom row) phase curves showing planets with homogeneous clouds and surfaces. The wavelength-dependent spectral features of desert, forest, and polar surfaces are taken from HAMSTER \citep{Roccetti2024} and modeled as Lambertian reflectors, while the ocean surface incorporates BRDF and BPDF treatments. Different columns refer to different wavelengths ($\lambda$): 500, 700, 900~nm.}
    \label{fig:phase_curves_hom_clouds}
\end{figure*}

For the same homogeneous planets, we simulate spectra and phase curves assuming a homogeneous cloud cover of 46\% (see Sec. \ref{sec:hom_planets}). Fig. \ref{fig:spectra_hom_clouds} presents the reflected (first row) and polarized (second row) light spectra for these cloudy planets. In reflected light, we observe a general increase in brightness compared to the cloud-free simulations (Sec. \ref{subsec:hom_planets}), with the effect being particularly pronounced for the ocean planet (dark blue). The polar planet (cyan), which already exhibits a very high surface albedo, is less affected by the presence of clouds. The addition of clouds also impacts the forest planet (green), where the VRE feature becomes less prominent in reflectance due to an increase in the continuum level before 750~nm. However, the effect is significantly stronger in polarization. With a cloudy atmosphere, the overall degree of polarization decreases substantially. For instance, in the ocean planet case, polarization at $\alpha$ = 90\degr drops from more than 80\% in the cloud-free scenario to between 5\% and 40\% in the cloudy case. The presence of clouds, due to multiple scattering, steepens the slope of the polarized spectra for the forest and desert planets while inverting the slopes for the ocean and Earth-like planet configurations. This highlights the superior diagnostic power of polarization compared to reflectance alone, as it becomes easier to distinguish between cloud-free and cloudy spectra. Moreover, the VRE feature is affected in polarization, as the characteristic drop in polarization between 700 and 800~nm is reduced.\\
\noindent A similar trend is observed in the phase curves for both reflected and polarized light (Fig. \ref{fig:phase_curves_hom_clouds}). In reflected light, the overall brightness increases slightly, particularly for the ocean planet, and additional features appear around $\alpha$ = 40\degr, corresponding to the cloudbow feature. In polarization, we again observe a significant reduction in the degree of linear polarization, yet new cloud-related features emerge compared to the cloud-free case (Fig. \ref{fig:phase_curves_hom}). The cloudbow is especially prominent in polarization and carries valuable information about cloud droplet microphysical properties, including size, composition, and shape \citep{emde2017, sterzik2020}. Additional polarization features also appear at large phase angles ($\alpha$ = 120\degr and 160\degr for $\lambda$ = 900~nm), which are associated with the change of the polarization direction.

\subsection{Models of increasing complexity}
\label{sec:homogeneous_planets_results}

\begin{figure*}[h]
    \centering
    \includegraphics[width=1\linewidth]{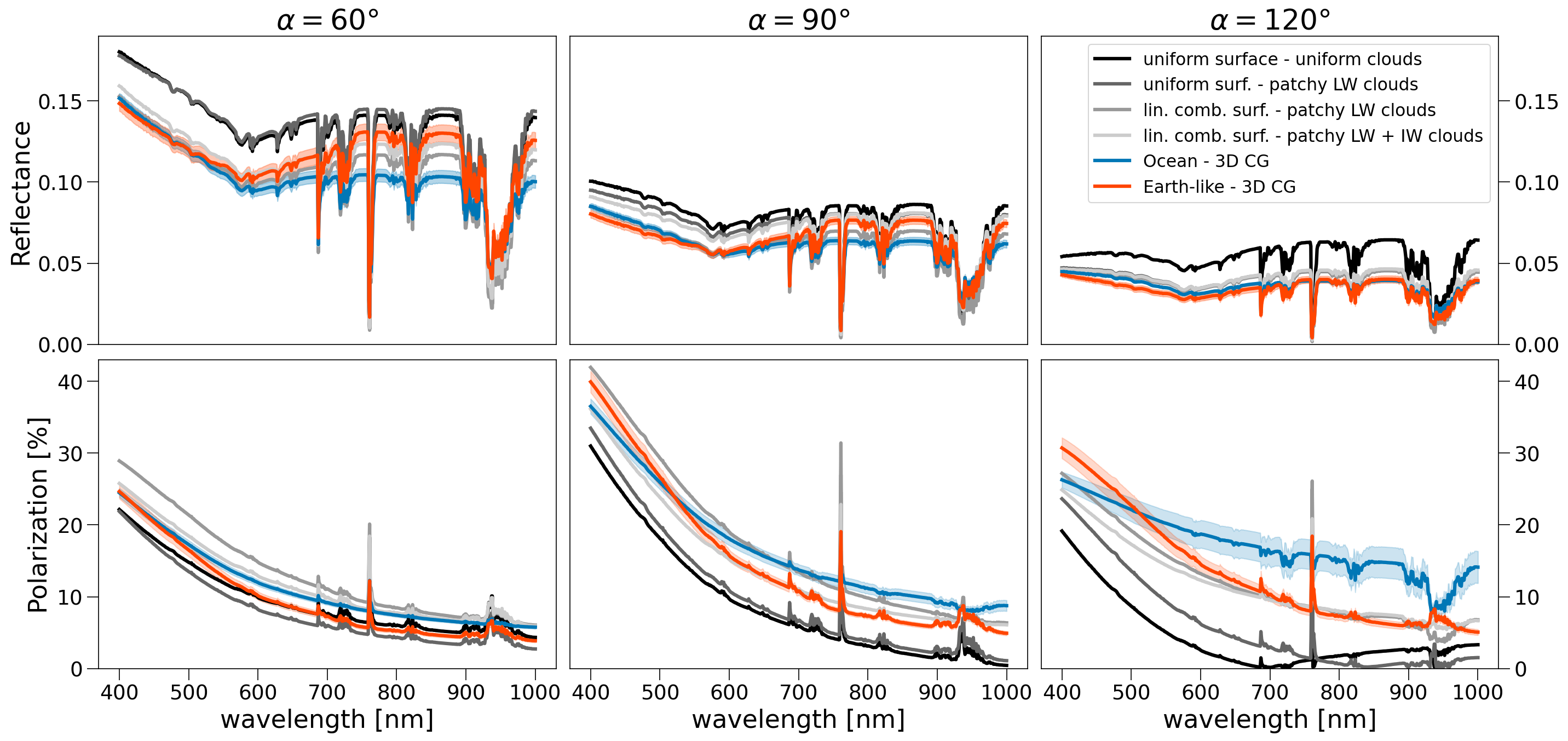}
    \caption{Comparison among spectra in reflected (first row) and polarized light (second row) of models of different complexity, from fully homogenous to complex Earth as an exoplanet simulation. Different columns refer to different phase angles ($\alpha$): 60, 90, 120\degr.}
    \label{fig:spectra_comparison}
\end{figure*}

\begin{figure*}[h]
    \centering
    \includegraphics[width=1\linewidth]{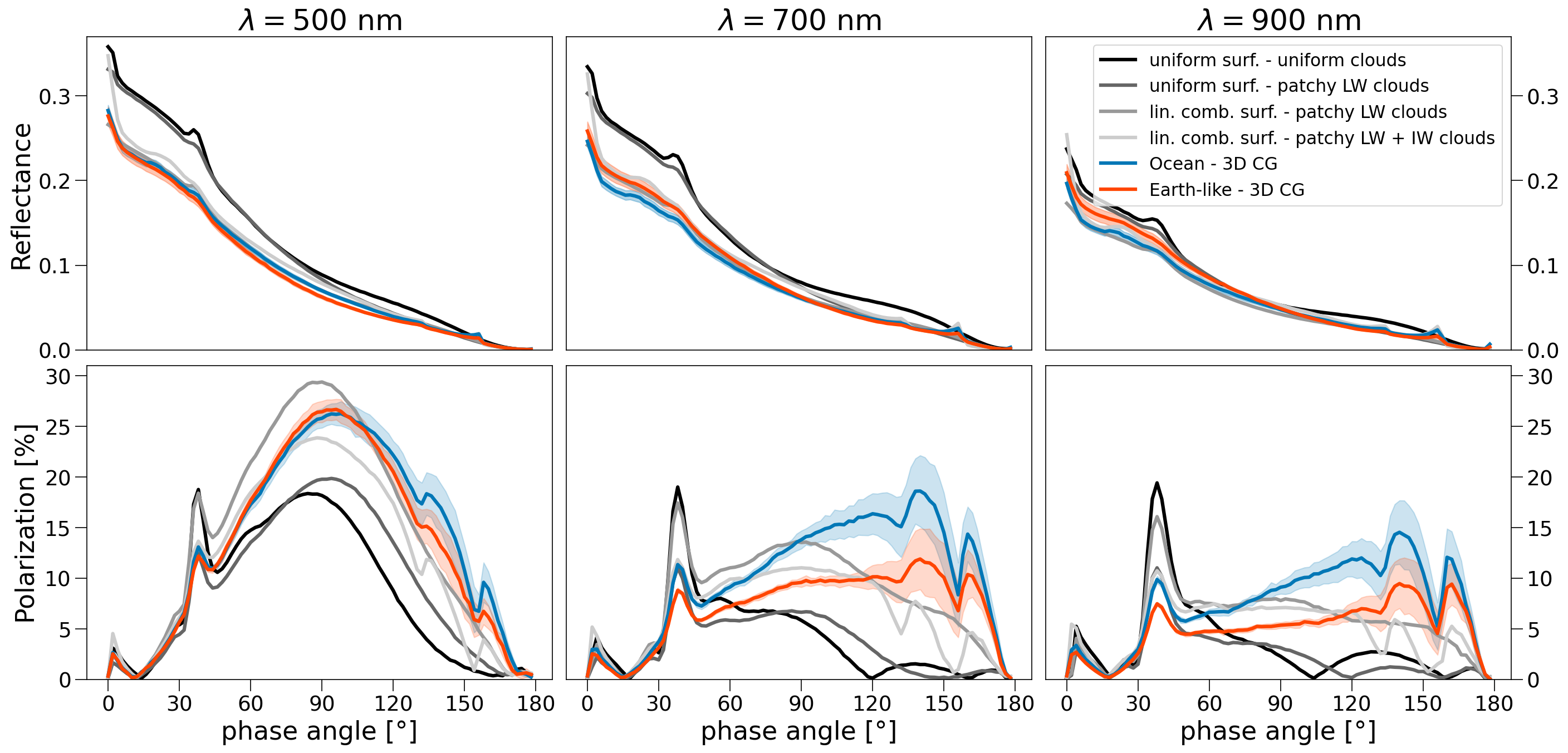}
    \caption{Reflected light (first row) and polarized light (second row) phase curves showing the influence of models of different complexities, from fully homogeneous to more complex Earth as an exoplanet simulation. Different columns refer to different wavelengths ($\lambda$): 500, 700, 900~nm.}
    \label{fig:phase_comparison}
\end{figure*}

Building on our simulations of homogeneous planets, we now investigate the impact of introducing inhomogeneities and increasing model complexity on reflectance and polarization. \cite{karalidi2012} demonstrated that models with horizontal inhomogeneities leave distinct traces in the polarization phase function and affect both the absolute values of reflectance and polarization in Earth-like exoplanet simulations. In this sensitivity study, we transition from homogeneous surface and cloud models to fully realistic Earth-like and ocean planet simulations.\\
\noindent In Fig. \ref{fig:spectra_comparison} we show the reflected and polarized light differences in the spectra due to the models of varying complexity. First, we notice the significant spread between the Ocean and Earth-like scenario already shown in \cite{Roccetti2025a}, way beyond the 1$\sigma$ cloud variability spread in the models (shaded areas) computed using various cloud fields from the ERA5 reanalysis product. This is particularly evident for $\alpha$ = 60\degr in reflectance, while the difference gets larger for $\alpha$ = 120\degr in polarization, where we observe a different behaviour both in the spectral slopes of the models, their continuum in the near-infrared (NIR) and in the behaviour of the spectral lines. Due to the presence (Ocean scenario) and absence (Earth-like scenario) of the ocean glint feature, we see a different behaviour of the water bands around 950~nm, as they are shown in absorption (spectral lines below the continuum) for the Ocean planet and in emission (spectral lines above the continuum) for the Earth-like case. This effect is already present in the $\alpha$ = 90\degr case, but gets enhanced at larger phase angles. For the uniform clouds and surface model (black line), we find it to substantially overestimate the reflectance of the planet and underestimate its polarization, as expected by a uniform cloud layer, where photons are reflected above the cloud deck. The effect gets more pronounced at large phase angles. Making the clouds more patchy (uniform surface - patchy LW clouds, dark grey line) shows improvement in the comparison with the more complex and realistic simulations in reflected light and at large phase angles. \\
\noindent We now change the uniform surface with a linear combination of surfaces with patchy LW clouds (light grey line). We find a much better correspondance with the Earth-like scenario in reflected light at $\alpha$ = 60\degr, where the spectral slope more closely matches the realistic Earth-like case, both in the Rayleigh scattering region and beyond the VRE. This improvement is due to the wavelength-dependent linear combination of surface types, which provides a more accurate representation of surface albedo. As a last improvement, we add IW clouds on top of the patchy LW clouds over the linear combination of surfaces (silver line). In reflected light, we notice a slight increase in the reflectance for small phase angles. \\
\noindent However, in polarization, we observe a notable difference between the simpler and more complex models, particularly at $\alpha$ = 90 and 120\degr. At $\alpha$ = 90\degr, uniform surfaces reduce the level of polarization compared to the linear combination of surfaces. This effect becomes even more pronounced at $\alpha$ = 120\degr, where polarization drops to zero just before 700~nm due to changes in the direction of the Stokes vectors when using a uniform surface. Additionally, the degree of linear polarization is significantly lower than in the Earth-like and Ocean planet scenarios, and the spectral slope is steeper than in the more realistic models. When introducing a linear combination of wavelength-dependent surfaces, the level of polarization becomes more comparable to the Earth-like scenario. This finding suggests that polarization is more sensitive to planetary surface features than reflectance alone, particularly at large phase angles. Accurately modeling wavelength-dependent surfaces is essential for properly interpreting disk-integrated spectra of exoplanets. However, while this approach improves polarization estimates, the slope in the Rayleigh scattering region and the direction of the water absorption bands still do not fully match the realistic model. This indicates that horizontally patchy clouds alone are insufficient to accurately model polarization, especially at high phase angles. These results suggest that capturing inhomogeneities in both cloud and surface modeling will be crucial for interpreting observations from the next generation of telescopes. Moreover, polarization provides stronger diagnostic capabilities for distinguishing different planetary scenarios and resolving potential retrieval degeneracies.\\
\noindent In Fig. \ref{fig:phase_comparison}, we analyze the impact of increasing model complexity on reflected and polarized phase curves at three different wavelengths: $\lambda$ = 500, 700, and 900~nm. Significant differences between the Ocean and Earth-like scenarios emerge primarily in polarization, particularly at 700 and 900~nm. We observe an increase in polarization due to ocean glint, which exceeds the variability introduced by cloud properties in our models, as discussed in \cite{Roccetti2025a}. In reflected light, uniform surface simulations consistently overestimate planetary reflectance, especially at shorter wavelengths. This discrepancy is even more pronounced in polarization, affecting the polarization peak associated with Rayleigh scattering (around $\alpha$ = 90\degr) at 700 and 900~nm. Moving from uniform clouds to horizontally patchy LW clouds on a uniform surface alters both the phase angle at which maximum polarization occurs and the prominence of the cloudbow feature. When introducing a linear combination of surfaces, we find improved agreement in both reflected and polarized light phase curves. However, the cloudbow feature remains substantially overestimated compared to the realistic 3D CG model without IW clouds. Additionally, including IW clouds generates polarization features at large phase angles ($\alpha$ = 138 and 158\degr), which result from ice crystal scattering properties, as explained in \cite{emde2017}. Overall, for $\lambda$ = 700 and 900~nm, simplified polarized phase curves fail to reproduce the benchmark Earth-like and Ocean models. These findings underscore the importance of accurately modeling 3D cloud inhomogeneities and subgrid variability when interpreting exoplanet phase curves, particularly in polarization. Our results suggest that distinguishing surface and atmospheric features using reflected light alone is more challenging than with polarization. However, in polarization, a homogeneous treatment of clouds and surface properties has a stronger impact on observational interpretation. This further emphasizes that polarization provides deeper insights into cloud properties than reflected light alone.\\
\noindent Since the next generation of telescopes and instruments, such as ANDES, will require long integration times to characterize rocky exoplanets, we examine in Appendix \ref{sec:appendix_A} the effect of time-averaging evolving cloud patterns over a typical 8-hour night of observation. In Figs. \ref{fig:spectra_smeared} and \ref{fig:phase_smeared}, we present the resulting impact on the spectra and phase curves for an Ocean planet under different cloud scenarios: a uniform liquid water cloud layer (with cloud properties as in Sec. \ref{sec:models_definition}), the 3D CG model with a 1$\sigma$ spread over 12 months, and an averaged model based on eight consecutive simulations, each using a distinct ERA5 cloud field as input to the 3D CG algorithm over an 8-hour period. This latter setup is designed to mimic the observation of an exoplanet over a single night, accounting for evolving weather patterns and changing scenery due to planet rotation. Our results show that even when cloud patterns are averaged over long integration times, the model predictions differ significantly from those based on a uniform cloud layer, and closely resemble the results from the 3D CG 1$\sigma$ spread model. This is because, even over extended integration periods, the instantaneous cloud distribution imprints its patchy structure on the reflected light. As a result, the ocean glint remains visible through cloud gaps at all times, and each timestep contributes to enhanced reflectance at small phase angles and increased polarization at large phase angles. Therefore, the resulting spectra and phase curves are not equivalent to those produced by a homogeneous, thinner cloud model.

\subsection{Absorption lines in high spectral resolution}

\begin{figure*}
    \centering
    \includegraphics[width=1.0\linewidth]{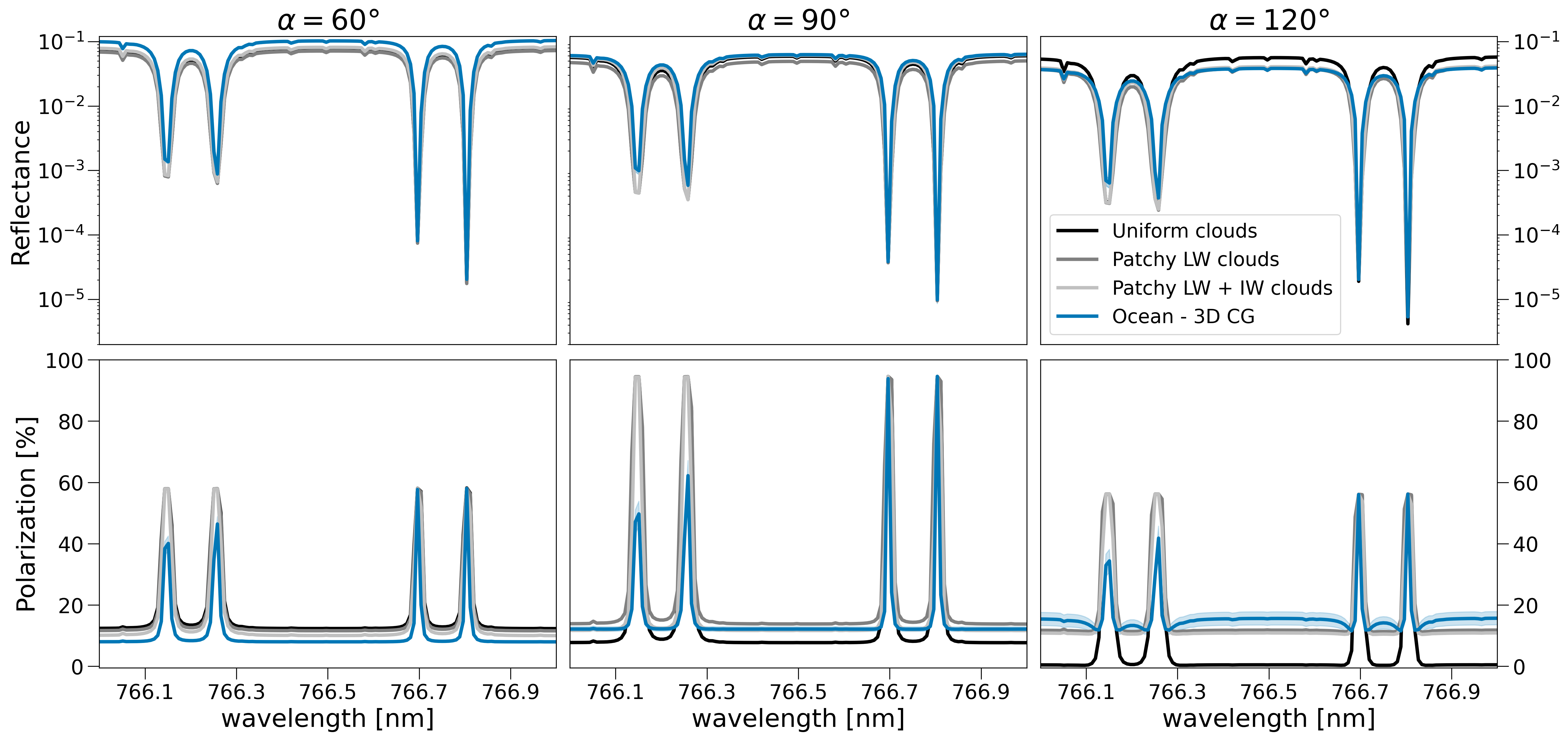}
    \caption{O$_2$-A band in reflected (top row) and polarized light (bottom row) at a spectral resolution of R = 100\,000. The absorption lines are modeled for an Ocean planet with three cloud treatments: uniform, homogeneous, and 3D CG clouds. Different columns refer to different phase angles ($\alpha$): 60, 90, 120\degr.}
    \label{fig:O2A}
\end{figure*}

\begin{figure*}
    \centering
    \includegraphics[width=1.0\linewidth]{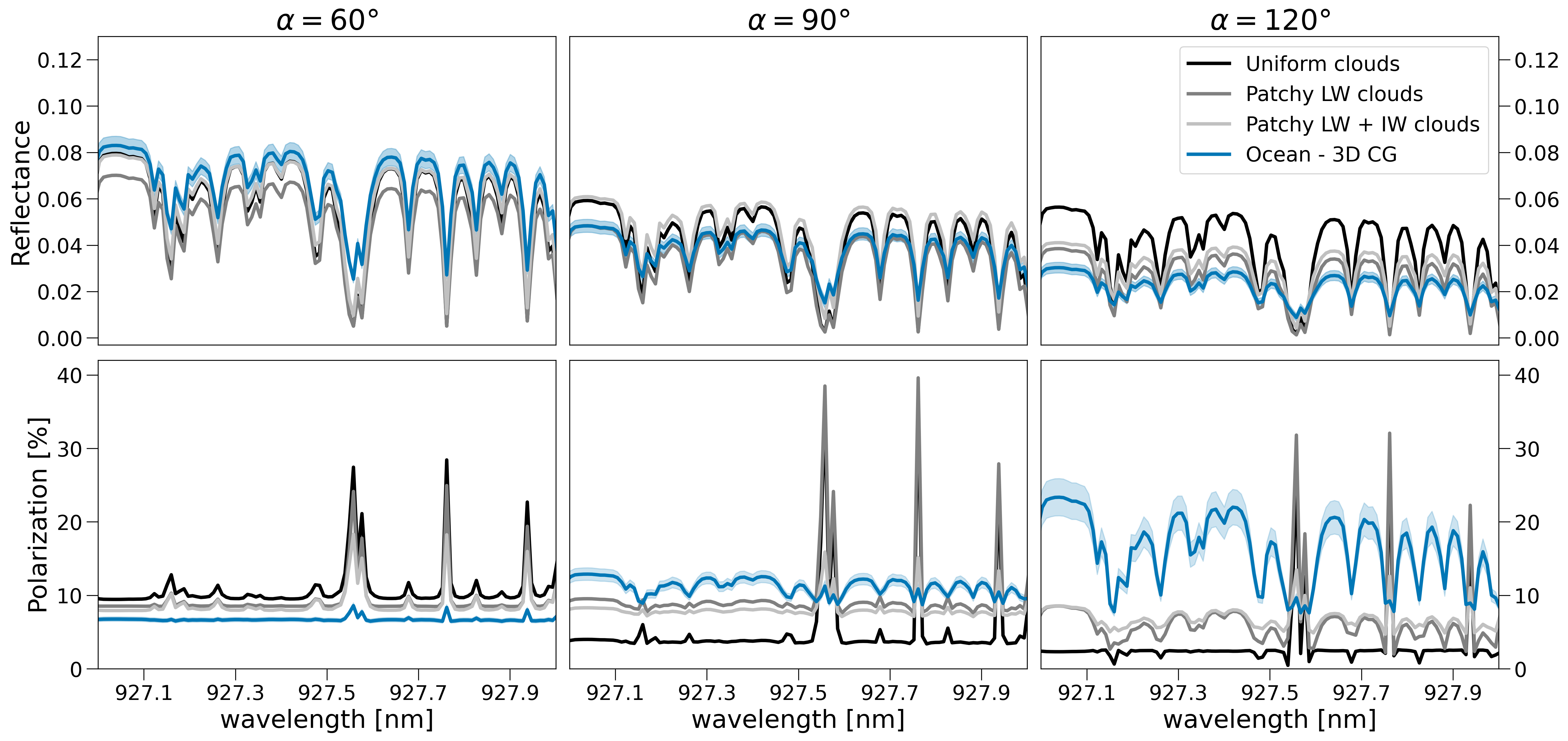}
    \caption{H$_2$O absorption lines in the Y band in reflected (top row) and polarized light (bottom row) at a spectral resolution of R = 100\,000. The absorption lines are modeled for an Ocean planet with three cloud treatments: uniform, homogeneous, and 3D CG clouds. Different columns refer to different phase angles ($\alpha$): 60, 90, 120\degr.}
    \label{fig:H2O_clouds}
\end{figure*}

\begin{figure*}
    \centering
    \includegraphics[width=1.0\linewidth]{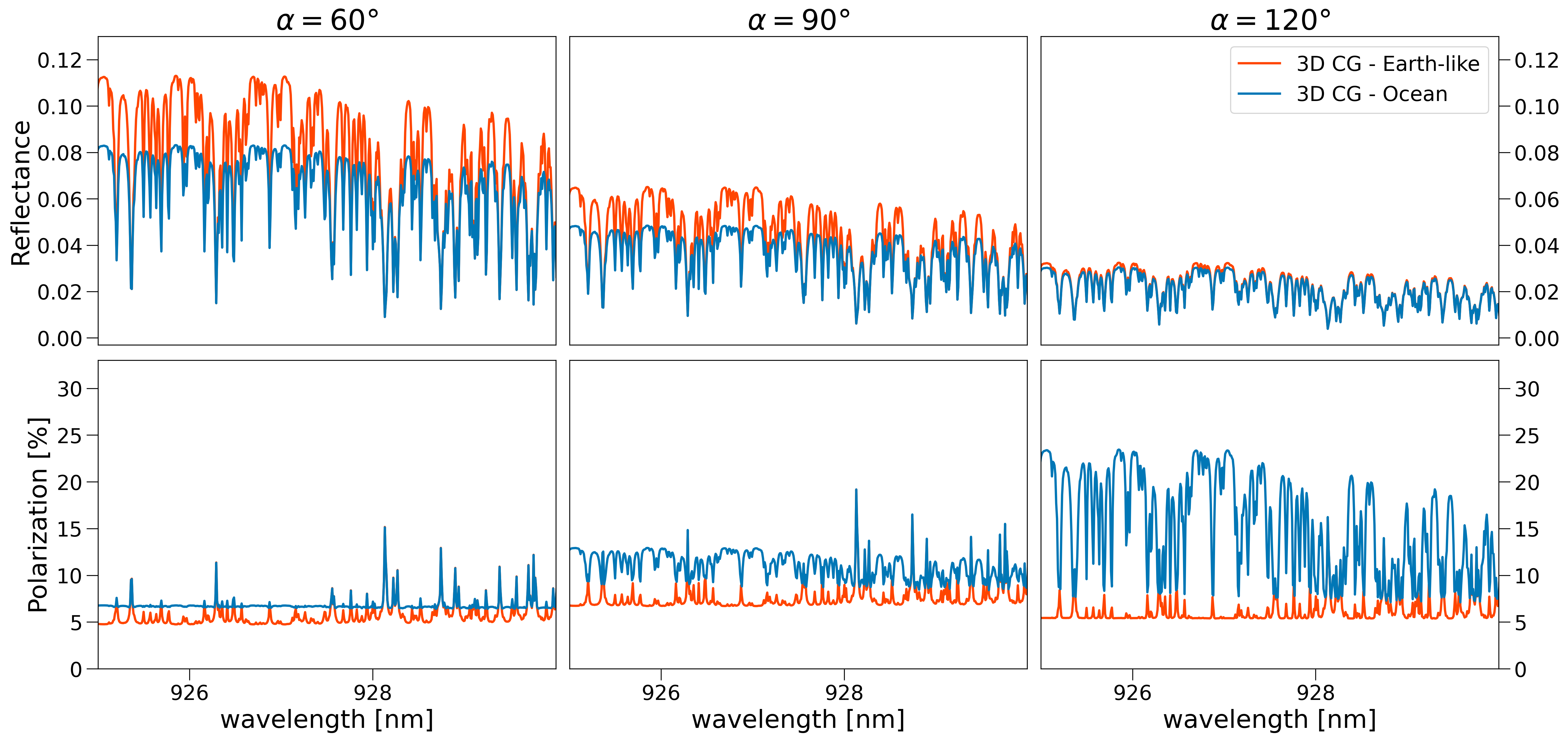}
    \caption{H$_2$O absorption lines in the Y band in reflected (top row) and polarized light (bottom row) at a spectral resolution of R = 100\,000. The absorption lines are modeled for an Ocean and an Earth-like planet scenario with 3D CG clouds. Different columns refer to different phase angles ($\alpha$): 60, 90, 120\degr.}
    \label{fig:H2O_surface}
\end{figure*}

We extensively discussed the influence of cloud and surface modeling approaches on the continuum levels of reflected and polarized light spectra. Now, we investigate the impact of models of varying complexity, ranging from uniform to patchy cloud modeling and different surface types, on absorption lines. Specifically, we focus on two molecular species: the O$_2$-A band around 780~nm and the H$_2$O lines in the Y band (920–950~nm). We perform high-resolution simulations with a spectral resolution of 100\,000, matching the expected capabilities of upcoming instruments on the ELT, such as ANDES.\\
\noindent Figure \ref{fig:O2A} illustrates the effect of different cloud models (uniform, patchy LW clouds, patchy LW and IW clouds and 3D CG simulations) over an ocean surface. In reflected light (please note the log scale), while the continuum is influenced by cloud modeling, the depth of the O$_2$-A line forest remains unaffected. In polarization, however, we observe changes not only in the continuum, particularly at $\alpha$ = 120\degr, but also in the depth of the absorption lines, which appear in emission. Notably, within the O$_2$-A band, the more saturated lines in reflectance remain unaffected by the choice of cloud model, while a more realistic treatment of clouds with the 3D CG lowers the polarization level of the emission lines, which are less saturated. This suggests that the conventional approach of retrieving cloud deck height using O$_2$-A line depth \citep{stam2008} is influenced by the presence of patchy clouds on a global scale.\\
\noindent We conduct the same comparison between uniform, patchy LW clouds, patchy LW and IW clouds and 3D CG models for the H$_2$O lines in the Y band, as shown in Fig. \ref{fig:H2O_clouds}. Here, we observe that uniform and patchy clouds affect the continuum, increasing the continuum level in reflected light and lowering it in polarization compared to the 3D CG ground truth models at $\alpha$ = 90 and 120\degr. Additionally, we identify an interesting behavior in the spectral lines. In reflected light, discrepancies appear in the line depths, while in polarization, the water lines are seen in absorption relative to the continuum in the 3D CG model, while they appear in emission for the simplistic uniform and patchy cloud approaches. The addition of the IW clouds does not have a large impact on the absorption lines.\\
\noindent In Fig. \ref{fig:H2O_surface}, we assess the impact of different surface types on water lines. The 3D CG Ocean and Earth-like models correspond to the spectra in Fig. \ref{fig:spectra_comparison}, but here we use a spectral resolution of 100\,000, focusing on the water band region. We observe that while the continuum is influenced by surface type, the water bands are shown in absorption for an ocean surface, while they appear in emission for a dry surface. This occurs because, in the Earth-like scenario, ocean glint is obscured by continents at large phase angles (see \citealt{Roccetti2025a}). This effect becomes more pronounced at larger phase angles, with the line depth remaining nearly consistent across all molecular lines both in absorption and emission.\\
\noindent This behavior was previously discussed by \cite{trees2022} for low spectral resolution simulations and homogeneous models. Here, we confirm that it persists even in more sophisticated simulations that incorporate sub-grid cloud inhomogeneities within the 3D CG framework. Furthermore, we demonstrate that the choice of cloud simulation approach also influences this line behavior in polarization. Therefore, accurately modeling clouds and their complex 3D structure is crucial for potential observations and for reliably extracting surface information from water lines.

\section{Contrast estimates for the ANDES golden sample}

\begin{figure*}
    \centering
    \includegraphics[width=1.0\linewidth]{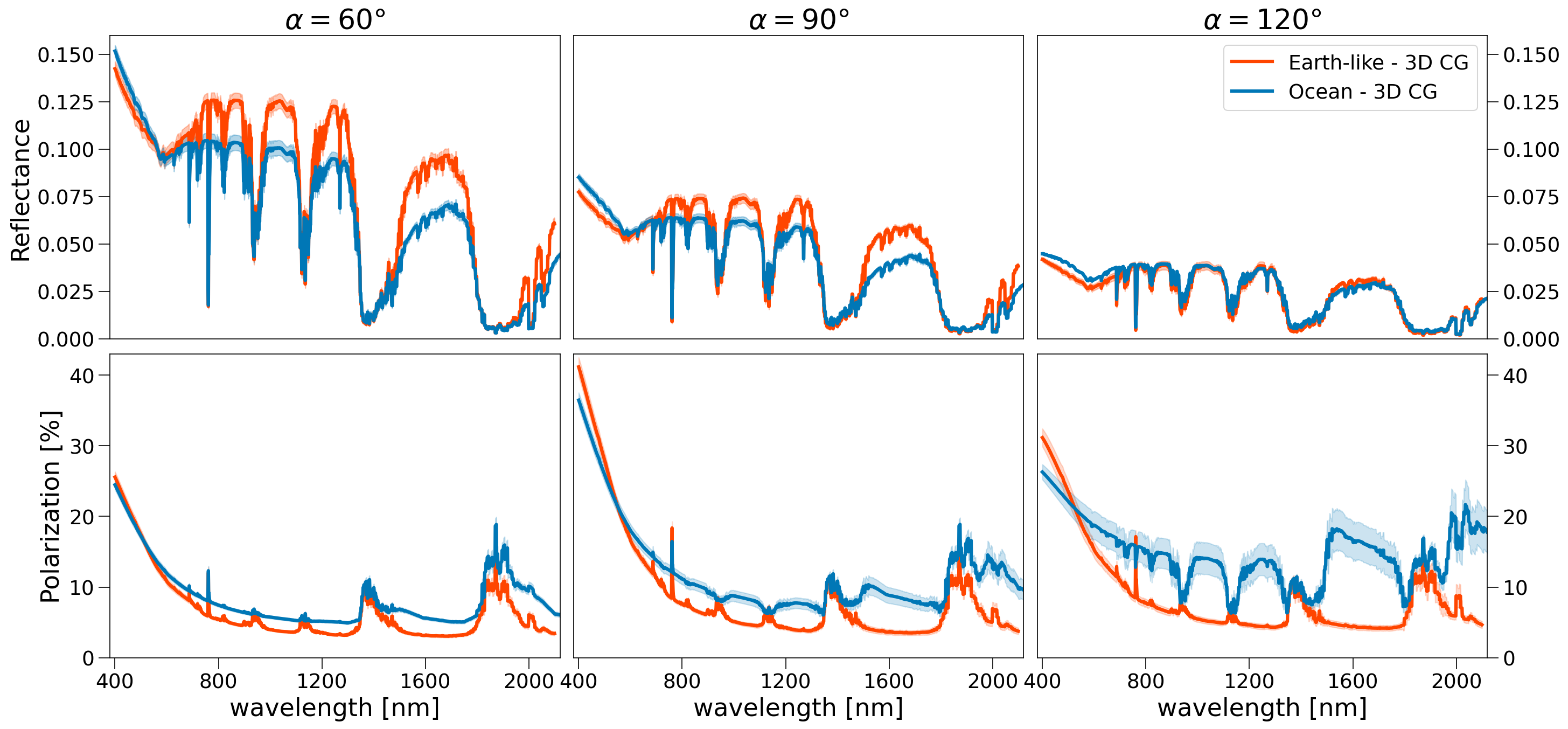}
    \caption{Comparison among spectra in reflected (first row) and polarized light (second row) of the Ocean and Earth-like planet scenarios for a M-dwarf star, for a wavelength range from 400 to 2500~nm. Different columns refer to different phase angles ($\alpha$): 60, 90, 120\degr.}
    \label{fig:M_dwarf}
\end{figure*}

In Fig. \ref{fig:M_dwarf}, we extend the ground-truth models for the Ocean and Earth-like planet scenarios from \cite{Roccetti2025a} into the NIR, up to 2500~nm. The figure illustrates that in reflected light, these two scenarios can be distinguished at small phase angles, where the Earth-like planet appears more reflective due to the high albedo of deserts, which is particularly significant in the NIR. In polarization, the distinction between the two scenarios becomes apparent at $\alpha$ = 90\degr, where water lines appear in absorption or emission depending on the underlying surface. This effect becomes even more pronounced at larger phase angles, further enhancing the diagnostic potential of polarized light. \\
\noindent With the enhanced cloud and surface modeling techniques presented in this work, we also provide updated flux contrast estimates for both intensity and polarization to support the preparation of future ground-based instruments, such as ANDES and PCS on the ELT, and space-based mission concepts like HWO. Using the scale factors for nearby exoplanets provided in Table~\ref{tab:scale_factor}, we refine the contrast estimates from previous studies (e.g., \citealt{palle2023}). It is important to note that rocky exoplanets orbiting M dwarfs are almost always tidally locked. Previous studies (e.g., \citealt{Way2017, Kopparapu2017}) have shown that tidal locking can result in atmospheric circulation and cloud patterns that differ significantly from those of Earth. In contrast, rocky exoplanets orbiting G-type stars, such as those targeted by HWO, are not expected to be tidally locked, and thus may exhibit more Earth-like cloud cover and distribution. \\
\noindent \cite{palle2023} presented a golden sample of rocky exoplanets around nearby M dwarfs to be characterized within a few observing nights with ANDES at the ELT. The reflected light contrast between the planet and the star is calculated using the maximum projected planet-star distance (i.e., when the planet is at $\alpha$ = 90\degr) and assuming an Earth-like albedo of 0.3. We now have the capabilities to provide improved reflected light contrast for this golden sample of exoplanets, assuming both an Earth-like and Ocean planet scenarios. In Table \ref{tab:contrasts} we provide our estimate of the contrast for the ANDES golden sample, with the addition of the newly discovered Barnard b exoplanet and a potential planet located at 1~AU orbiting Alpha Cen A. \\
\begin{table*}[h]
    \caption{\label{tab:contrasts} Refined contrast estimates for rocky exoplanets. We compare the ANDES golden sample with a potential Earth-like and Ocean planet around Alpha Centauri A. We compare the contrast estimates from \cite{palle2023} with our updated calculations using the 3D Cloud Generator (3D CG) in both reflected and polarized light. Additionally, we present contrast values obtained for an Ocean planet with a coarser horizontal grid resolution (zoom-out x100) for comparison.}
    \centering
    \resizebox{0.89\textwidth}{!}{%
    \begin{tabular}{c|cc|c|c|cc}
    \toprule 
    & \multicolumn{2}{c|}{$\text{C}_{\text{flux}} [\text{3D CG}]$} & $\text{C}_{\text{flux}} [\text{zoom-out}]$ & $\text{C}_{\text{flux}} [\text{\cite{palle2023}}]$ & \multicolumn{2}{c}{$\text{C}_{\text{pol}} [\text{3D CG}]$} \\  
    Name & Ocean & Earth-like & Ocean & & Ocean & Earth-like\\
    \hline
    Proxima Cen b & 5.69 $\cdot 10^{-8}$  & 5.83 $\cdot 10^{-8}$  & 9.01 $\cdot 10^{-8}$   & 11.2 $\cdot 10^{-8}$ & 1.15 $\cdot 10^{-8}$  & 1.11 $\cdot 10^{-8}$\\
    Ross 128 b    & 6.23 $\cdot 10^{-8}$  & 6.45 $\cdot 10^{-8}$  & 9.97 $\cdot 10^{-8}$   & 12.5 $\cdot 10^{-8}$ & 1.28 $\cdot 10^{-8}$  & 1.23 $\cdot 10^{-8}$\\
    GJ 273 b      & 3.78 $\cdot 10^{-8}$  & 3.87 $\cdot 10^{-8}$  & 5.98 $\cdot 10^{-8}$   & 7.52 $\cdot 10^{-8}$ & 0.77$\cdot 10^{-8}$  & 0.74$\cdot 10^{-8}$\\
    Wolf 1061 c   & 4.81 $\cdot 10^{-8}$  & 4.93 $\cdot 10^{-8}$  & 7.62 $\cdot 10^{-8}$   & 9.57 $\cdot 10^{-8}$ & 0.97$\cdot 10^{-8}$  & 0.94$\cdot 10^{-8}$\\
    GJ 682 c      & 8.09 $\cdot 10^{-8}$  & 8.29 $\cdot 10^{-8}$  & 12.8 $\cdot 10^{-8}$  & 16.0 $\cdot 10^{-8}$ & 1.64 $\cdot 10^{-8}$  & 1.58 $\cdot 10^{-8}$\\
    Barnard b     & 12.1 $\cdot 10^{-8}$  & 12.4 $\cdot 10^{-8}$  & 19.1 $\cdot 10^{-8}$  &                      & 2.44 $\cdot 10^{-8}$  & 2.35 $\cdot 10^{-8}$\\
    Alpha Cen A   & 1.17 $\cdot 10^{-10}$ & 1.19 $\cdot 10^{-10}$ & 1.85 $\cdot 10^{-10}$  &                      & 0.24$\cdot 10^{-10}$ & 0.23$\cdot 10^{-10}$\\
    \bottomrule
    \end{tabular}
    }
\end{table*}
\noindent We find that our contrasts are lower than the ones obtained by \cite{palle2023} by a factor of two, and that the flux and polarization contrasts between Earth-like and Ocean planet scenarios are remarkably similar. For example, for Proxima b with a phase angle of 90\degr, \cite{palle2023} estimated a contrast of 11.2 $\cdot 10^{-8}$, while our calculated value is 5.8 $\cdot 10^{-8}$ for an Earth-like planet. This discrepancy arises from differences in reflectance estimation. \cite{palle2023} used planetary albedo estimates from \cite{turbet2016}, who simulated various possible climates and atmospheric states for Proxima b using a General Circulation Model (GCM). However, GCMs operate at relatively coarse grid resolutions, and when coupled with a 3D radiation scheme, they simulate planets with a lower spatial resolution than our approach.\\
\noindent The impact of spatial resolution on radiative transfer calculations has been previously highlighted by \cite{robinson2011}, who demonstrated that a minimum resolution of 100 pixels was necessary to achieve acceptable fits to EPOXI spacecraft data of Earth as an exoplanet. In \cite{Roccetti2025a}, we further explore how planetary reflectance varies with horizontal resolution. Specifically, when comparing our results to those of \cite{turbet2016}, who employed a 64 × 48 grid (closer to our zoom-out x100 case), we find that our reflected light contrasts closely match theirs. Moreover, when applying the same zoom-out x100 resolution to cloud modeling, our contrast estimates align closely with those from \cite{palle2023}. This underscores the crucial role of horizontal resolution in obtaining accurate contrast predictions and reconciling model outputs with observations. Additionally, we find that Barnard b exhibits a higher contrast compared to the original five exoplanets in the golden sample presented in \cite{palle2023}, reaching 1.2 $\cdot 10^{-7}$.\\
\noindent We also provide contrast estimates in polarization, which are approximately one-fifth of the flux contrast obtained with the 3D CG. While the lower contrast significantly impacts the planet’s detectability and characterization, polarization offers key advantages. It is largely unaffected by telluric contamination and enhances star-planet separation, as F-, G-, and K-type stars are expected to emit almost entirely unpolarized light. Moreover, when using a coronagraph, polarimetric techniques help suppress stellar speckles, as demonstrated with ZIMPOL \citep{hunziker2020}, which is particularly beneficial at small angular separations \citep{beuzit2019}.

\section{Discussion and conclusions}
\label{sec:conclusions}

In this work, we build upon the improved cloud and surface modeling presented in \cite{Roccetti2025a} to assess the importance of detailed cloud and surface properties in studying rocky exoplanets. Using the 3D radiative transfer model MYSTIC, with the same horizontal and vertical resolutions, we analyze how an improved treatment of clouds and surfaces affects reflected and polarized light spectra and phase curves. Additionally, our approach enables a comparison of the insights gained from combining spectroscopy and spectropolarimetry versus intensity-alone measurements for future observations of rocky exoplanets with ANDES and PCS at the ELT, as well as the mission concept HWO.\\
\noindent We compare the ground-truth models presented in \cite{Roccetti2025a} on reflected and polarized light spectra and phase curves to those of homogeneous planet models and models of increasing complexity. Our analysis leads to several key findings:
\begin{enumerate}
    \item Polarization provides stronger diagnostic capabilities than intensity alone in distinguishing between cloud-free and cloudy exoplanets. The spectral slope and polarization fraction are highly sensitive to clouds, while the cloudbow feature offers valuable insights into cloud microphysical properties. Reflectance loses diagnostic power as phase angles increase, while polarization shows the opposite trend.
    \item A uniform surface fails to reproduce both reflected and polarized spectra and phase curves. Incorporating a linear combination of wavelength-dependent surface types significantly improves agreement with ground-truth spectra. Polarization spectra are particularly sensitive to surface properties, especially at large phase angles, both in the continuum and in water absorption lines. 
    \item Simplified cloud treatments, such as homogeneous cloud models and single-layer clouds with averaged properties, introduce significant inaccuracies, even when attempting to mimic the effects of long observational averaging. Polarization phase curves are more sensitive to cloud properties, particularly through the cloudbow and ice crystals features, making them a crucial tool for cloud characterization.
    \item Water absorption lines in polarization appear in absorption when ocean glint is present and in emission for dry planets, as previously reported by \cite{trees2022} for homogeneous models. This behavior persists even at high spectral resolution (R = 100\,000), confirming the potential of water lines as surface diagnostics. However, simplistic cloud models can alter the appearance of water lines, potentially affecting the interpretation of ocean detection on exoplanets.
    \item Using our ground-truth reflected and polarized light models for an Earth-like scenario, we calculate contrast estimates for the ANDES golden sample \citep{palle2023}, including Barnard b and a hypothetical Earth-like planet at 1~AU from Alpha Centauri A. Compared to previous studies \citep{turbet2016, palle2023}, our results show that reflected light contrast estimates are overestimated by a factor of two when using coarse horizontal resolution and simplified cloud and surface models. Additionally, we provide contrast estimates in polarization, which are approximately one-fifth of the reflected light flux contrast.
\end{enumerate}
\noindent These findings strongly suggest that retrieval frameworks for reflected-light observations of rocky exoplanets should account for wavelength-dependent surface albedo properties and patchy cloud models. As shown by \cite{Wang2022}, neglecting wavelength-dependent variations in surface albedo in retrieval frameworks can lead to substantially biased estimates of atmospheric and cloud properties. Notably, we demonstrate that a linear combination of just four surface types (ocean, desert, forest, and polar regions) achieves good agreement with complex ground-truth models, highlighting a practical approach for future retrievals. Additionally, our results reinforce the potential impact of polarization in exoplanet characterization. By combining polarization with intensity-only spectroscopy, we can enhance diagnostic capabilities, reduce retrieval degeneracies, and improve the characterization of surface and atmospheric properties. Additionally, finer spatial grid resolutions for radiative transfer calculations should be considered to avoid biases when comparing models with observations.\\
\noindent Future studies should assess the feasibility of polarized-light observations with next-generation telescopes, determining whether sufficient contrast can be achieved for robust exoplanet characterization. More broadly, our results demonstrate that homogeneous models fail to accurately represent Earth as an exoplanet, emphasizing the need for more advanced modeling approaches in the exoplanet community. Clouds play a crucial role in shaping observables, making their accurate treatment essential for reliable simulations. Both simplistic cloud treatments and low horizontal resolution in radiative transfer models significantly impact the accuracy of simulations, reinforcing the need for high-resolution, state-of-the-art models for meaningful comparisons with observations.\\
\noindent Ultimately, we demonstrate that polarization is a powerful tool for characterizing rocky exoplanets, distinguishing between different planetary scenarios both in the visible and NIR, and providing deeper insights into their physical and chemical properties. As future telescopes and missions become operational, incorporating these advanced modeling techniques will be crucial for interpreting observations and understanding the diversity of rocky exoplanets. Furthermore, our refined contrast estimates play a key role in instrument design and in precisely determining integration times.

\section{Data Availability}
The spectra and phase curve data from this study are publicly available through a Jupyter notebook at \url{https://github.com/giulia-roccetti/Earth_as_an_exoplanet_Part_II}.

\begin{acknowledgements}
GR and JVS were supported by the Munich Institute for Astro-, Particle and BioPhysics (MIAPbP) which is funded by the Deutsche Forschungsgemeinschaft (DFG, German Research Foundation) under Germany´s Excellence Strategy – EXC-2094 – 390783311.
\end{acknowledgements}

\bibliographystyle{aa} 
\bibliography{bibliography}

\begin{thebibliography}{73}
\expandafter\ifx\csname natexlab\endcsname\relax\def\natexlab#1{#1}\fi

\bibitem[{{Alei} {et~al.}(2024){Alei}, {Quanz}, {Konrad}, {Garvin}, {Kofman}, {Mandell}, {Angerhausen}, {Molli{\`e}re}, {Meyer}, {Robinson}, {Rugheimer}, \& {the LIFE Collaboration}}]{alei2024}
{Alei}, E., {Quanz}, S.~P., {Konrad}, B.~S., {et~al.} 2024, arXiv e-prints, arXiv:2406.13037

\bibitem[{{Anderson} {et~al.}(1986){Anderson}, {Clough}, {Kneizys}, {Chetwynd}, \& {Shettle}}]{anderson1986}
{Anderson}, G.~P., {Clough}, S.~A., {Kneizys}, F.~X., {Chetwynd}, J.~H., \& {Shettle}, E.~P. 1986, {AFGL atmospheric constituent profiles (0.120km)}

\bibitem[{{Beuzit} {et~al.}(2019){Beuzit}, {Vigan}, {Mouillet}, {Dohlen}, {Gratton}, {Boccaletti}, {Sauvage}, {Schmid}, {Langlois}, {Petit}, {Baruffolo}, {Feldt}, {Milli}, {Wahhaj}, {Abe}, {Anselmi}, {Antichi}, {Barette}, {Baudrand}, {Baudoz}, {Bazzon}, {Bernardi}, {Blanchard}, {Brast}, {Bruno}, {Buey}, {Carbillet}, {Carle}, {Cascone}, {Chapron}, {Charton}, {Chauvin}, {Claudi}, {Costille}, {De Caprio}, {de Boer}, {Delboulb{\'e}}, {Desidera}, {Dominik}, {Downing}, {Dupuis}, {Fabron}, {Fantinel}, {Farisato}, {Feautrier}, {Fedrigo}, {Fusco}, {Gigan}, {Ginski}, {Girard}, {Giro}, {Gisler}, {Gluck}, {Gry}, {Henning}, {Hubin}, {Hugot}, {Incorvaia}, {Jaquet}, {Kasper}, {Lagadec}, {Lagrange}, {Le Coroller}, {Le Mignant}, {Le Ruyet}, {Lessio}, {Lizon}, {Llored}, {Lundin}, {Madec}, {Magnard}, {Marteaud}, {Martinez}, {Maurel}, {M{\'e}nard}, {Mesa}, {M{\"o}ller-Nilsson}, {Moulin}, {Moutou}, {Orign{\'e}}, {Parisot}, {Pavlov}, {Perret}, {Pragt}, {Puget}, {Rabou}, {Ramos}, {Reess}, {Rigal}, {Rochat}, {Roelfsema}, {Rousset},
  {Roux}, {Saisse}, {Salasnich}, {Santambrogio}, {Scuderi}, {Segransan}, {Sevin}, {Siebenmorgen}, {Soenke}, {Stadler}, {Suarez}, {Tiph{\`e}ne}, {Turatto}, {Udry}, {Vakili}, {Waters}, {Weber}, {Wildi}, {Zins}, \& {Zurlo}}]{beuzit2019}
{Beuzit}, J.~L., {Vigan}, A., {Mouillet}, D., {et~al.} 2019, \aap, 631, A155

\bibitem[{{Borsa} {et~al.}(2019){Borsa}, {Rainer}, {Bonomo}, {Barbato}, {Fossati}, {Malavolta}, {Nascimbeni}, {Lanza}, {Esposito}, {Affer}, {Andreuzzi}, {Benatti}, {Biazzo}, {Bignamini}, {Brogi}, {Carleo}, {Claudi}, {Cosentino}, {Covino}, {Damasso}, {Desidera}, {Garrido Rubio}, {Giacobbe}, {Gonz{\'a}lez-{\'A}lvarez}, {Harutyunyan}, {Knapic}, {Leto}, {Ligi}, {Maggio}, {Maldonado}, {Mancini}, {Fiorenzano}, {Masiero}, {Micela}, {Molinari}, {Pagano}, {Pedani}, {Piotto}, {Pino}, {Poretti}, {Scandariato}, {Smareglia}, \& {Sozzetti}}]{borsa2019}
{Borsa}, F., {Rainer}, M., {Bonomo}, A.~S., {et~al.} 2019, \aap, 631, A34

\bibitem[{Buehler {et~al.}(2005)Buehler, Eriksson, Kuhn, {von Engeln}, \& Verdes}]{buehler2005}
Buehler, S., Eriksson, P., Kuhn, T., {von Engeln}, A., \& Verdes, C. 2005, Journal of Quantitative Spectroscopy and Radiative Transfer, 91, 65

\bibitem[{{Buenzli} \& {Schmid}(2009)}]{Buenzli&Schmid2009A}
{Buenzli}, E. \& {Schmid}, H.~M. 2009, \aap, 504, 259

\bibitem[{{Bugatti} {et~al.}(2024){Bugatti}, {Lovis}, {Pepe}, {Blind}, {Billot}, {Chazelas}, \& {Turbet}}]{Bugatti2024}
{Bugatti}, M., {Lovis}, C., {Pepe}, F., {et~al.} 2024, arXiv e-prints, arXiv:2412.20879

\bibitem[{{Buras} \& {Mayer}(2011)}]{buas2011}
{Buras}, R. \& {Mayer}, B. 2011, \jqsrt, 112, 434

\bibitem[{{Cadieux} {et~al.}(2024){Cadieux}, {Doyon}, {MacDonald}, {Turbet}, {Artigau}, {Lim}, {Radica}, {Fauchez}, {Salhi}, {Dang}, {Albert}, {Coulombe}, {Cowan}, {Lafreni{\`e}re}, {L'Heureux}, {Piaulet-Ghorayeb}, {Benneke}, {Cloutier}, {Charnay}, {Cook}, {Fournier-Tondreau}, {Plotnykov}, \& {Valencia}}]{cadieux2024}
{Cadieux}, C., {Doyon}, R., {MacDonald}, R.~J., {et~al.} 2024, \apjl, 970, L2

\bibitem[{{Charbonneau} {et~al.}(1999){Charbonneau}, {Noyes}, {Korzennik}, {Nisenson}, {Jha}, {Vogt}, \& {Kibrick}}]{charbonneau1999}
{Charbonneau}, D., {Noyes}, R.~W., {Korzennik}, S.~G., {et~al.} 1999, \apjl, 522, L145

\bibitem[{{Collier Cameron} {et~al.}(1999){Collier Cameron}, {Horne}, {Penny}, \& {James}}]{collier-cameron1999}
{Collier Cameron}, A., {Horne}, K., {Penny}, A., \& {James}, D. 1999, \nat, 402, 751

\bibitem[{{Collier Cameron} {et~al.}(2002){Collier Cameron}, {Horne}, {Penny}, \& {Leigh}}]{collier-cameron2002}
{Collier Cameron}, A., {Horne}, K., {Penny}, A., \& {Leigh}, C. 2002, \mnras, 330, 187

\bibitem[{{Costa Silva} {et~al.}(2024){Costa Silva}, {Demangeon}, {Santos}, {Ehrenreich}, {Lovis}, {Chakraborty}, {Lendl}, {Pepe}, {Cristiani}, {Rebolo}, {Zapatero-Osorio}, {Adibekyan}, {Alibert}, {Allart}, {Allende Prieto}, {Azevedo Silva}, {Borsa}, {Bourrier}, {Cristo}, {Di Marcantonio}, {Esparza-Borges}, {Figueira}, {Gonz{\'a}lez Hern{\'a}ndez}, {Herrero-Cisneros}, {Lo Curto}, {Martins}, {Mehner}, {Nunes}, {Palle}, {Pelletier}, {Seidel}, {Silva}, {Sousa}, {Sozzetti}, {Steiner}, {Su{\'a}rez Mascare{\~n}o}, \& {Udry}}]{CostaSilva2024}
{Costa Silva}, A.~R., {Demangeon}, O.~D.~S., {Santos}, N.~C., {et~al.} 2024, \aap, 689, A8

\bibitem[{{Cotton} {et~al.}(2017){Cotton}, {Marshall}, {Bailey}, {Kedziora-Chudczer}, {Bott}, {Marsden}, \& {Carter}}]{cotton2017}
{Cotton}, D.~V., {Marshall}, J.~P., {Bailey}, J., {et~al.} 2017, \mnras, 467, 873

\bibitem[{{Dong} {et~al.}(2018){Dong}, {Jin}, {Lingam}, {Airapetian}, {Ma}, \& {van der Holst}}]{dong2018}
{Dong}, C., {Jin}, M., {Lingam}, M., {et~al.} 2018, Proceedings of the National Academy of Science, 115, 260

\bibitem[{{Edwards} \& {Tinetti}(2022)}]{Edwards&Tinetti2022}
{Edwards}, B. \& {Tinetti}, G. 2022, \aj, 164, 15

\bibitem[{{Emde} {et~al.}(2011){Emde}, {Buras}, \& {Mayer}}]{emde2011}
{Emde}, C., {Buras}, R., \& {Mayer}, B. 2011, \jqsrt, 112, 1622

\bibitem[{Emde {et~al.}(2016)Emde, Buras-Schnell, Kylling, Mayer, Gasteiger, Hamann, Kylling, Richter, Pause, Dowling, \& Bugliaro}]{emde2016}
Emde, C., Buras-Schnell, R., Kylling, A., {et~al.} 2016, Geoscientific Model Development, 9, 1647

\bibitem[{{Emde} {et~al.}(2017){Emde}, {Buras-Schnell}, {Sterzik}, \& {Bagnulo}}]{emde2017}
{Emde}, C., {Buras-Schnell}, R., {Sterzik}, M., \& {Bagnulo}, S. 2017, \aap, 605, A2

\bibitem[{Eriksson {et~al.}(2011)Eriksson, Buehler, Davis, Emde, \& Lemke}]{eriksson2011}
Eriksson, P., Buehler, S., Davis, C., Emde, C., \& Lemke, O. 2011, Journal of Quantitative Spectroscopy and Radiative Transfer, 112, 1551

\bibitem[{{Garc{\'\i}a Mu{\~n}oz}(2015)}]{munoz2015}
{Garc{\'\i}a Mu{\~n}oz}, A. 2015, International Journal of Astrobiology, 14, 379

\bibitem[{Gasteiger {et~al.}(2014)Gasteiger, Emde, Mayer, Buras, Buehler, \& Lemke}]{gasteiger2014}
Gasteiger, J., Emde, C., Mayer, B., {et~al.} 2014, Journal of Quantitative Spectroscopy and Radiative Transfer, 148, 99

\bibitem[{{Gonz{\'a}lez Hern{\'a}ndez} {et~al.}(2024){Gonz{\'a}lez Hern{\'a}ndez}, {Su{\'a}rez Mascare{\~n}o}, {Silva}, {Stefanov}, {Faria}, {Tabernero}, {Sozzetti}, {Rebolo}, {Pepe}, {Santos}, {Cristiani}, {Lovis}, {Dumusque}, {Figueira}, {Lillo-Box}, {Nari}, {Benatti}, {Hobson}, {Castro-Gonz{\'a}lez}, {Allart}, {Passegger}, {Zapatero Osorio}, {Adibekyan}, {Alibert}, {Allende Prieto}, {Bouchy}, {Damasso}, {D'Odorico}, {Di Marcantonio}, {Ehrenreich}, {Lo Curto}, {Santos}, {Martins}, {Mehner}, {Micela}, {Molaro}, {Nunes}, {Palle}, {Sousa}, \& {Udry}}]{Gonazalez-Hernandez2024}
{Gonz{\'a}lez Hern{\'a}ndez}, J.~I., {Su{\'a}rez Mascare{\~n}o}, A., {Silva}, A.~M., {et~al.} 2024, \aap, 690, A79

\bibitem[{{Gordon} {et~al.}(2023){Gordon}, {Karalidi}, {Bott}, {Miles-P{\'a}ez}, {Mulder}, \& {Stam}}]{gordon2023}
{Gordon}, K.~E., {Karalidi}, T., {Bott}, K.~M., {et~al.} 2023, \apj, 945, 166

\bibitem[{{Greene} {et~al.}(2023){Greene}, {Bell}, {Ducrot}, {Dyrek}, {Lagage}, \& {Fortney}}]{greene2023}
{Greene}, T.~P., {Bell}, T.~J., {Ducrot}, E., {et~al.} 2023, \nat, 618, 39

\bibitem[{Hall {et~al.}(2018)Hall, Thompson, Handley, \& Queloz}]{hall2018}
Hall, R.~D., Thompson, S.~J., Handley, W., \& Queloz, D. 2018, Monthly Notices of the Royal Astronomical Society, 479, 2968

\bibitem[{{Hersbach} {et~al.}(2020){Hersbach}, {Bell}, {Berrisford}, {Hirahara}, {Hor{\'a}nyi}, {Mu{\~n}oz-Sabater}, {Nicolas}, {Peubey}, {Radu}, {Schepers}, {Simmons}, {Soci}, {Abdalla}, {Abellan}, {Balsamo}, {Bechtold}, {Biavati}, {Bidlot}, {Bonavita}, {De Chiara}, {Dahlgren}, {Dee}, {Diamantakis}, {Dragani}, {Flemming}, {Forbes}, {Fuentes}, {Geer}, {Haimberger}, {Healy}, {Hogan}, {H{\'o}lm}, {Janiskov{\'a}}, {Keeley}, {Laloyaux}, {Lopez}, {Lupu}, {Radnoti}, {de Rosnay}, {Rozum}, {Vamborg}, {Villaume}, \& {Th{\'e}paut}}]{Hersbach2020}
{Hersbach}, H., {Bell}, B., {Berrisford}, P., {et~al.} 2020, Quarterly Journal of the Royal Meteorological Society, 146, 1999

\bibitem[{{Hoeijmakers} {et~al.}(2018){Hoeijmakers}, {Snellen}, \& {van Terwisga}}]{hoeijmakers2018}
{Hoeijmakers}, H.~J., {Snellen}, I.~A.~G., \& {van Terwisga}, S.~E. 2018, \aap, 610, A47

\bibitem[{{Hu} {et~al.}(2024){Hu}, {Bello-Arufe}, {Zhang}, {Paragas}, {Zilinskas}, {van Buchem}, {Bess}, {Patel}, {Ito}, {Damiano}, {Scheucher}, {Oza}, {Knutson}, {Miguel}, {Dragomir}, {Brandeker}, \& {Demory}}]{hu2024}
{Hu}, R., {Bello-Arufe}, A., {Zhang}, M., {et~al.} 2024, \nat, 630, 609

\bibitem[{{Hunziker} {et~al.}(2020){Hunziker}, {Schmid}, {Mouillet}, {Milli}, {Zurlo}, {Delorme}, {Abe}, {Avenhaus}, {Baruffolo}, {Bazzon}, {Boccaletti}, {Baudoz}, {Beuzit}, {Carbillet}, {Chauvin}, {Claudi}, {Costille}, {Daban}, {Desidera}, {Dohlen}, {Dominik}, {Downing}, {Engler}, {Feldt}, {Fusco}, {Ginski}, {Gisler}, {Girard}, {Gratton}, {Henning}, {Hubin}, {Kasper}, {Keller}, {Langlois}, {Lagadec}, {Martinez}, {Maire}, {Menard}, {Meyer}, {Pavlov}, {Pragt}, {Puget}, {Quanz}, {Rickman}, {Roelfsema}, {Salasnich}, {Sauvage}, {Siebenmorgen}, {Sissa}, {Snik}, {Suarez}, {Szul{\'a}gyi}, {Thalmann}, {Turatto}, {Udry}, {van Holstein}, {Vigan}, \& {Wildi}}]{hunziker2020}
{Hunziker}, S., {Schmid}, H.~M., {Mouillet}, D., {et~al.} 2020, \aap, 634, A69

\bibitem[{{Karalidi} \& {Stam}(2012)}]{karalidi2012}
{Karalidi}, T. \& {Stam}, D.~M. 2012, \aap, 546, A56

\bibitem[{{Kasper} {et~al.}(2021){Kasper}, {Cerpa Urra}, {Pathak}, {Bonse}, {Nousiainen}, {Engler}, {Heritier}, {Kammerer}, {Leveratto}, {Rajani}, {Bristow}, {Le Louarn}, {Madec}, {Str{\"o}bele}, {Verinaud}, {Glauser}, {Quanz}, {Helin}, {Keller}, {Snik}, {Boccaletti}, {Chauvin}, {Mouillet}, {Kulcs{\'a}r}, \& {Raynaud}}]{kasper2021}
{Kasper}, M., {Cerpa Urra}, N., {Pathak}, P., {et~al.} 2021, The Messenger, 182, 38

\bibitem[{{Kasting} {et~al.}(1993){Kasting}, {Whitmire}, \& {Reynolds}}]{kasting1993}
{Kasting}, J.~F., {Whitmire}, D.~P., \& {Reynolds}, R.~T. 1993, \icarus, 101, 108

\bibitem[{{Kemp} {et~al.}(1987){Kemp}, {Henson}, {Steiner}, \& {Powell}}]{kemp1987}
{Kemp}, J.~C., {Henson}, G.~D., {Steiner}, C.~T., \& {Powell}, E.~R. 1987, \nat, 326, 270

\bibitem[{{Kopparapu} {et~al.}(2013){Kopparapu}, {Ramirez}, {Kasting}, {Eymet}, {Robinson}, {Mahadevan}, {Terrien}, {Domagal-Goldman}, {Meadows}, \& {Deshpande}}]{Kopparapu2013}
{Kopparapu}, R.~K., {Ramirez}, R., {Kasting}, J.~F., {et~al.} 2013, \apj, 765, 131

\bibitem[{{Kopparapu} {et~al.}(2017){Kopparapu}, {Wolf}, {Arney}, {Batalha}, {Haqq-Misra}, {Grimm}, \& {Heng}}]{Kopparapu2017}
{Kopparapu}, R.~k., {Wolf}, E.~T., {Arney}, G., {et~al.} 2017, \apj, 845, 5

\bibitem[{{Lovis} {et~al.}(2022){Lovis}, {Blind}, {Chazelas}, {K{\"u}hn}, {Genolet}, {Hughes}, {Sordet}, {Schnell}, {Turbet}, {Fusco}, {Sauvage}, {Bugatti}, {Billot}, {Hagelberg}, {Hocini}, \& {Guyon}}]{Lovis2022}
{Lovis}, C., {Blind}, N., {Chazelas}, B., {et~al.} 2022, in Society of Photo-Optical Instrumentation Engineers (SPIE) Conference Series, Vol. 12184, Ground-based and Airborne Instrumentation for Astronomy IX, ed. C.~J. {Evans}, J.~J. {Bryant}, \& K.~{Motohara}, 121841Q

\bibitem[{{Luger} \& {Barnes}(2015)}]{luger&barnes2015}
{Luger}, R. \& {Barnes}, R. 2015, Astrobiology, 15, 119

\bibitem[{{Martins} {et~al.}(2015){Martins}, {Santos}, {Figueira}, {Faria}, {Montalto}, {Boisse}, {Ehrenreich}, {Lovis}, {Mayor}, {Melo}, {Pepe}, {Sousa}, {Udry}, \& {Cunha}}]{martins2015}
{Martins}, J.~H.~C., {Santos}, N.~C., {Figueira}, P., {et~al.} 2015, \aap, 576, A134

\bibitem[{{Mayer}(2009)}]{mayer2009}
{Mayer}, B. 2009, in European Physical Journal Web of Conferences, Vol.~1, European Physical Journal Web of Conferences, 75

\bibitem[{Mayer \& Kylling(2005)}]{mayer2005}
Mayer, B. \& Kylling, A. 2005, Atmospheric Chemistry and Physics, 5, 1855

\bibitem[{{Monta{\~n}{\'e}s-Rodr{\'\i}guez} {et~al.}(2006){Monta{\~n}{\'e}s-Rodr{\'\i}guez}, {Pall{\'e}}, {Goode}, \& {Mart{\'\i}n-Torres}}]{Montanes-Rodriguez2006}
{Monta{\~n}{\'e}s-Rodr{\'\i}guez}, P., {Pall{\'e}}, E., {Goode}, P.~R., \& {Mart{\'\i}n-Torres}, F.~J. 2006, \apj, 651, 544

\bibitem[{{National Academies of Sciences} \& Medicine(2021)}]{US_decadal_survey}
{National Academies of Sciences}, E. \& Medicine. 2021, {Pathways to Discovery in Astronomy and Astrophysics for the 2020s}

\bibitem[{{Nortmann} {et~al.}(2025){Nortmann}, {Lesjak}, {Yan}, {Cont}, {Czesla}, {Lavail}, {Rains}, {Nagel}, {Boldt-Christmas}, {Hatzes}, {Reiners}, {Piskunov}, {Kochukhov}, {Heiter}, {Shulyak}, {Rengel}, \& {Seemann}}]{Nortmann2025}
{Nortmann}, L., {Lesjak}, F., {Yan}, F., {et~al.} 2025, \aap, 693, A213

\bibitem[{{Pall{\'e}} {et~al.}(2023){Pall{\'e}}, {Biazzo}, {Bolmont}, {Molliere}, {Poppenhaeger}, {Birkby}, {Brogi}, {Chauvin}, {Chiavassa}, {Hoeijmakers}, {Lellouch}, {Lovis}, {Maiolino}, {Nortmann}, {Parviainen}, {Pino}, {Turbet}, {Wender}, {Albrecht}, {Antoniucci}, {Barros}, {Beaudoin}, {Benneke}, {Boisse}, {Bonomo}, {Borsa}, {Brandeker}, {Brandner}, {Buchhave}, {Cheffot}, {Deborde}, {Debras}, {Doyon}, {Di Marcantonio}, {Giacobbe}, {Gonzalez Hernandez}, {Helled}, {Kreidberg}, {Machado}, {Maldonado}, {Marconi}, {Canto Martins}, {Miceli}, {Mordasini}, {N'Diaye}, {Niedzielski}, {Nisini}, {Origlia}, {Peroux}, {Pietrow}, {Pinna}, {Rauscher}, {Reffert}, {Rousselot}, {Sanna}, {Simonnin}, {Suarez Mascareno}, {Zanutta}, \& {Zechmeister}}]{palle2023}
{Pall{\'e}}, E., {Biazzo}, K., {Bolmont}, E., {et~al.} 2023, arXiv e-prints, arXiv:2311.17075

\bibitem[{{Pino} {et~al.}(2020){Pino}, {D{\'e}sert}, {Brogi}, {Malavolta}, {Wyttenbach}, {Line}, {Hoeijmakers}, {Fossati}, {Bonomo}, {Nascimbeni}, {Panwar}, {Affer}, {Benatti}, {Biazzo}, {Bignamini}, {Borsa}, {Carleo}, {Claudi}, {Cosentino}, {Covino}, {Damasso}, {Desidera}, {Giacobbe}, {Harutyunyan}, {Lanza}, {Leto}, {Maggio}, {Maldonado}, {Mancini}, {Micela}, {Molinari}, {Pagano}, {Piotto}, {Poretti}, {Rainer}, {Scandariato}, {Sozzetti}, {Allart}, {Borsato}, {Bruno}, {Di Fabrizio}, {Ehrenreich}, {Fiorenzano}, {Frustagli}, {Lavie}, {Lovis}, {Magazz{\`u}}, {Nardiello}, {Pedani}, \& {Smareglia}}]{Pino2020}
{Pino}, L., {D{\'e}sert}, J.-M., {Brogi}, M., {et~al.} 2020, \apjl, 894, L27

\bibitem[{{Prinoth} {et~al.}(2024{\natexlab{a}}){Prinoth}, {Hoeijmakers}, {Morris}, {Lam}, {Kitzmann}, {Sedaghati}, {Seidel}, {Lee}, {Thorsbro}, {Borsato}, {Damasceno}, {Pelletier}, \& {Seifahrt}}]{Prinoth2024b}
{Prinoth}, B., {Hoeijmakers}, H.~J., {Morris}, B.~M., {et~al.} 2024{\natexlab{a}}, \aap, 685, A60

\bibitem[{{Prinoth} {et~al.}(2024{\natexlab{b}}){Prinoth}, {Sedaghati}, {Seidel}, {Hoeijmakers}, {Brahm}, {Thorsbro}, \& {Jord{\'a}n}}]{Prinoth2024a}
{Prinoth}, B., {Sedaghati}, E., {Seidel}, J.~V., {et~al.} 2024{\natexlab{b}}, arXiv e-prints, arXiv:2406.08558

\bibitem[{{Quanz} {et~al.}(2022){Quanz}, {Ottiger}, {Fontanet}, {Kammerer}, {Menti}, {Dannert}, {Gheorghe}, {Absil}, {Airapetian}, {Alei}, {Allart}, {Angerhausen}, {Blumenthal}, {Buchhave}, {Cabrera}, {Carri{\'o}n-Gonz{\'a}lez}, {Chauvin}, {Danchi}, {Dandumont}, {Defr{\'e}re}, {Dorn}, {Ehrenreich}, {Ertel}, {Fridlund}, {Garc{\'\i}a Mu{\~n}oz}, {Gasc{\'o}n}, {Girard}, {Glauser}, {Grenfell}, {Guidi}, {Hagelberg}, {Helled}, {Ireland}, {Janson}, {Kopparapu}, {Korth}, {Kozakis}, {Kraus}, {L{\'e}ger}, {Leedj{\"a}rv}, {Lichtenberg}, {Lillo-Box}, {Linz}, {Liseau}, {Loicq}, {Mahendra}, {Malbet}, {Mathew}, {Mennesson}, {Meyer}, {Mishra}, {Molaverdikhani}, {Noack}, {Oza}, {Pall{\'e}}, {Parviainen}, {Quirrenbach}, {Rauer}, {Ribas}, {Rice}, {Romagnolo}, {Rugheimer}, {Schwieterman}, {Serabyn}, {Sharma}, {Stassun}, {Szul{\'a}gyi}, {Wang}, {Wunderlich}, {Wyatt}, \& {LIFE Collaboration}}]{Quanz2022}
{Quanz}, S.~P., {Ottiger}, M., {Fontanet}, E., {et~al.} 2022, \aap, 664, A21

\bibitem[{{Rauer} {et~al.}(2024){Rauer}, {Aerts}, {Cabrera}, {Deleuil}, {Erikson}, {Gizon}, {Goupil}, {Heras}, {Lorenzo-Alvarez}, {Marliani}, {Martin-Garcia}, {Mas-Hesse}, {O'Rourke}, {Osborn}, {Pagano}, {Piotto}, {Pollacco}, {Ragazzoni}, {Ramsay}, {Udry}, {Appourchaux}, {Benz}, {Brandeker}, {G{\"u}del}, {Janot-Pacheco}, {Kabath}, {Kjeldsen}, {Min}, {Santos}, {Smith}, {Suarez}, {Werner}, {Aboudan}, {Abreu}, {Acu a}, {Adams}, {Adibekyan}, {Affer}, {Agneray}, {Agnor}, {Aguirre B{\o}rsen-Koch}, {Ahmed}, {Aigrain}, {Al-Bahlawan}, {Alcacera Gil}, {Alei}, {Alencar}, {Alexander}, {Alfonso-Garz{\'o}n}, {Alibert}, {Allende Prieto}, {Almeida}, {Alonso Sobrino}, {Altavilla}, {Althaus}, {Alonso Alvarez Trujillo}, {Amarsi}, {Ammler-von Eiff}, {Am{\^o}res}, {Andrade}, {Antoniadis-Karnavas}, {Ant{\'o}nio}, {Aparicio del Moral}, {Appolloni}, {Arena}, {Armstrong}, {Aroca Aliaga}, {Asplund}, {Audenaert}, {Auricchio}, {Avelino}, {Baeke}, {Bailli{\'e}}, {Balado}, {Ballber Balaguer{\'o}}, {Balestra}, {Ball}, {Ballans}, {Ballot},
  {Barban}, {Barbary}, {Barbieri}, {Barcel{\'o} Forteza}, {Barker}, {Barklem}, {Barnes}, {Barrado Navascues}, {Barragan}, {Baruteau}, {Basu}, {Baudin}, {Baumeister}, {Bayliss}, {Bazot}, {Beck}, {Bedding}, {Belkacem}, {Bellinger}, {Benatti}, {Benomar}, {B{\'e}rard}, {Bergemann}, {Bergomi}, {Bernardo}, {Biazzo}, {Bignamini}, {Bigot}, {Billot}, {Binet}, {Biondi}, {Biondi}, {Birch}, {Bitsch}, {Bluhm Ceballos}, {B{\'o}di}, {Bogn{\'a}r}, {Boisse}, {Bolmont}, {Bonanno}, {Bonavita}, {Bonfanti}, {Bonfils}, {Bonito}, {Bonomo}, {B{\"o}rner}, {Boro Saikia}, {Borreguero Mart{\'\i}n}, {Borsa}, {Borsato}, {Bossini}, {Bouchy}, {Bou{\'e}}, {Boufleur}, {Boumier}, {Bourrier}, {Bowman}, {Bozzo}, {Bradley}, {Bray}, {Bressan}, {Breton}, {Brienza}, {Brito}, {Brogi}, {Brown}, {Brown}, {Brun}, {Bruno}, {Bruns}, {Buchhave}, {Bugnet}, {Buldgen}, {Burgess}, {Busatta}, {Busso}, {Buzasi}, {Caballero}, {Cabral}, {Cabrero Gomez}, {Calderone}, {Cameron}, {Cameron}, {Campante}, {Campos Gestal}, {Canto Martins}, {Cara}, {Carone}, {Carrasco},
  {Casagrande}, {Casewell}, {Cassisi}, {Castellani}, {Castro}, {Catala}, {Catal{\'a}n Fern{\'a}ndez}, {Catelan}, {Cegla}, {Cerruti}, {Cessa}, {Chadid}, {Chaplin}, {Charpinet}, {Chiappini}, {Chiarucci}, {Chiavassa}, {Chinellato}, {Chirulli}, {Christensen-Dalsgaard}, {Church}, {Claret}, {Clarke}, {Claudi}, {Clermont}, {Coelho}, {Coelho}, {Cogato}, {Colom{\'e}}, {Condamin}, {Conde Garc{\'\i}a}, \& {Conseil}}]{Rauer2024}
{Rauer}, H., {Aerts}, C., {Cabrera}, J., {et~al.} 2024, arXiv e-prints, arXiv:2406.05447

\bibitem[{{Robinson} {et~al.}(2011){Robinson}, {Meadows}, {Crisp}, {Deming}, {A'Hearn}, {Charbonneau}, {Livengood}, {Seager}, {Barry}, {Hearty}, {Hewagama}, {Lisse}, {McFadden}, \& {Wellnitz}}]{robinson2011}
{Robinson}, T.~D., {Meadows}, V.~S., {Crisp}, D., {et~al.} 2011, Astrobiology, 11, 393

\bibitem[{Roccetti {et~al.}(2024)Roccetti, Bugliaro, G\"odde, Emde, Hamann, Manev, Sterzik, \& Wehrum}]{Roccetti2024}
Roccetti, G., Bugliaro, L., G\"odde, F., {et~al.} 2024, Atmospheric Measurement Techniques, 17, 6025

\bibitem[{{Roccetti} {et~al.}(2025){Roccetti}, {Emde}, {Sterzik}, {Manev}, {Seidel}, \& {Bagnulo}}]{Roccetti2025a}
{Roccetti}, G., {Emde}, C., {Sterzik}, M.~F., {et~al.} 2025, \aap, 697, A170

\bibitem[{{Rodler} {et~al.}(2013){Rodler}, {K{\"u}rster}, {L{\'o}pez-Morales}, \& {Ribas}}]{rodler2013}
{Rodler}, F., {K{\"u}rster}, M., {L{\'o}pez-Morales}, M., \& {Ribas}, I. 2013, Astronomische Nachrichten, 334, 188

\bibitem[{{Scandariato} {et~al.}(2021){Scandariato}, {Borsa}, {Sicilia}, {Malavolta}, {Biazzo}, {Bonomo}, {Bruno}, {Claudi}, {Covino}, {Di Marcantonio}, {Esposito}, {Frustagli}, {Lanza}, {Maldonado}, {Maggio}, {Mancini}, {Micela}, {Nardiello}, {Rainer}, {Singh}, {Sozzetti}, {Affer}, {Benatti}, {Bignamini}, {Biliotti}, {Capuzzo-Dolcetta}, {Carleo}, {Cosentino}, {Damasso}, {Desidera}, {Garcia de Gurtubai}, {Ghedina}, {Giacobbe}, {Giani}, {Harutyunyan}, {Hernandez}, {Hernandez Diaz}, {Knapic}, {Leto}, {Mart{\'\i}nez Fiorenzano}, {Molinari}, {Nascimbeni}, {Pagano}, {Pedani}, {Piotto}, {Poretti}, \& {Stoev}}]{scandariato2021}
{Scandariato}, G., {Borsa}, F., {Sicilia}, D., {et~al.} 2021, \aap, 646, A159

\bibitem[{{Seidel} {et~al.}(2020{\natexlab{a}}){Seidel}, {Lendl}, {Bourrier}, {Ehrenreich}, {Allart}, {Sousa}, {Cegla}, {Bonfils}, {Conod}, {Grandjean}, {Wyttenbach}, {Astudillo-Defru}, {Bayliss}, {Heng}, {Lavie}, {Lovis}, {Melo}, {Pepe}, {S{\'e}gransan}, \& {Udry}}]{seidel2020b}
{Seidel}, J.~V., {Lendl}, M., {Bourrier}, V., {et~al.} 2020{\natexlab{a}}, \aap, 643, A45

\bibitem[{{Seidel} {et~al.}(2020{\natexlab{b}}){Seidel}, {Lendl}, {Bourrier}, {Ehrenreich}, {Allart}, {Sousa}, {Cegla}, {Bonfils}, {Conod}, {Grandjean}, {Wyttenbach}, {Astudillo-Defru}, {Bayliss}, {Heng}, {Lavie}, {Lovis}, {Melo}, {Pepe}, {S{\'e}gransan}, \& {Udry}}]{seidel2020c}
{Seidel}, J.~V., {Lendl}, M., {Bourrier}, V., {et~al.} 2020{\natexlab{b}}, \aap, 643, A45

\bibitem[{{Seidel} {et~al.}(2025){Seidel}, {Prinoth}, {Pino}, {dos Santos}, {Chakraborty}, {Parmentier}, {Sedaghati}, {Wardenier}, {Farret Jentink}, {Zapatero Osorio}, {Allart}, {Ehrenreich}, {Lendl}, {Roccetti}, {Damasceno}, {Bourrier}, {Lillo-Box}, {Hoeijmakers}, {Pall{\'e}}, {Santos}, {Su{\'a}rez Mascare{\~n}o}, {Sousa}, {Tabernero}, \& {Pepe}}]{Seidel2025}
{Seidel}, J.~V., {Prinoth}, B., {Pino}, L., {et~al.} 2025, \nat, 639, 902

\bibitem[{{Selsis} {et~al.}(2007){Selsis}, {Kasting}, {Levrard}, {Paillet}, {Ribas}, \& {Delfosse}}]{Selsis2007}
{Selsis}, F., {Kasting}, J.~F., {Levrard}, B., {et~al.} 2007, \aap, 476, 1373

\bibitem[{{Snellen} {et~al.}(2008){Snellen}, {Albrecht}, {de Mooij}, \& {Le Poole}}]{Snellen2008}
{Snellen}, I.~A.~G., {Albrecht}, S., {de Mooij}, E.~J.~W., \& {Le Poole}, R.~S. 2008, \aap, 487, 357

\bibitem[{{Spring} {et~al.}(2022){Spring}, {Birkby}, {Pino}, {Alonso}, {Hoyer}, {Young}, {Coelho}, {Nespral}, \& {L{\'o}pez-Morales}}]{spring2022}
{Spring}, E.~F., {Birkby}, J.~L., {Pino}, L., {et~al.} 2022, \aap, 659, A121

\bibitem[{{Stam}(2008)}]{stam2008}
{Stam}, D.~M. 2008, \aap, 482, 989

\bibitem[{{Sterzik} {et~al.}(2020){Sterzik}, {Bagnulo}, {Emde}, \& {Manev}}]{sterzik2020}
{Sterzik}, M.~F., {Bagnulo}, S., {Emde}, C., \& {Manev}, M. 2020, \aap, 639, A89

\bibitem[{{Tinetti} {et~al.}(2018){Tinetti}, {Drossart}, {Eccleston}, {Hartogh}, {Heske}, {Leconte}, {Micela}, {Ollivier}, {Pilbratt}, {Puig}, {Turrini}, {Vandenbussche}, {Wolkenberg}, {Beaulieu}, {Buchave}, {Ferus}, {Griffin}, {Guedel}, {Justtanont}, {Lagage}, {Machado}, {Malaguti}, {Min}, {N{\o}rgaard-Nielsen}, {Rataj}, {Ray}, {Ribas}, {Swain}, {Szabo}, {Werner}, {Barstow}, {Burleigh}, {Cho}, {Coud{\'e} du Foresto}, {Coustenis}, {Decin}, {Encrenaz}, {Galand}, {Gillon}, {Helled}, {Morales}, {Garc{\'\i}a Mu{\~n}oz}, {Moneti}, {Pagano}, {Pascale}, {Piccioni}, {Pinfield}, {Sarkar}, {Selsis}, {Tennyson}, {Triaud}, {Venot}, {Waldmann}, {Waltham}, {Wright}, {Amiaux}, {Augu{\`e}res}, {Berth{\'e}}, {Bezawada}, {Bishop}, {Bowles}, {Coffey}, {Colom{\'e}}, {Crook}, {Crouzet}, {Da Peppo}, {Sanz}, {Focardi}, {Frericks}, {Hunt}, {Kohley}, {Middleton}, {Morgante}, {Ottensamer}, {Pace}, {Pearson}, {Stamper}, {Symonds}, {Rengel}, {Renotte}, {Ade}, {Affer}, {Alard}, {Allard}, {Altieri}, {Andr{\'e}}, {Arena}, {Argyriou},
  {Aylward}, {Baccani}, {Bakos}, {Banaszkiewicz}, {Barlow}, {Batista}, {Bellucci}, {Benatti}, {Bernardi}, {B{\'e}zard}, {Blecka}, {Bolmont}, {Bonfond}, {Bonito}, {Bonomo}, {Brucato}, {Brun}, {Bryson}, {Bujwan}, {Casewell}, {Charnay}, {Pestellini}, {Chen}, {Ciaravella}, {Claudi}, {Cl{\'e}dassou}, {Damasso}, {Damiano}, {Danielski}, {Deroo}, {Di Giorgio}, {Dominik}, {Doublier}, {Doyle}, {Doyon}, {Drummond}, {Duong}, {Eales}, {Edwards}, {Farina}, {Flaccomio}, {Fletcher}, {Forget}, {Fossey}, {Fr{\"a}nz}, {Fujii}, {Garc{\'\i}a-Piquer}, {Gear}, {Geoffray}, {G{\'e}rard}, {Gesa}, {Gomez}, {Graczyk}, {Griffith}, {Grodent}, {Guarcello}, {Gustin}, {Hamano}, {Hargrave}, {Hello}, {Heng}, {Herrero}, {Hornstrup}, {Hubert}, {Ida}, {Ikoma}, {Iro}, {Irwin}, {Jarchow}, {Jaubert}, {Jones}, {Julien}, {Kameda}, {Kerschbaum}, {Kervella}, {Koskinen}, {Krijger}, {Krupp}, {Lafarga}, {Landini}, {Lellouch}, {Leto}, {Luntzer}, {Rank-L{\"u}ftinger}, {Maggio}, {Maldonado}, {Maillard}, {Mall}, {Marquette}, {Mathis}, {Maxted}, {Matsuo},
  {Medvedev}, {Miguel}, {Minier}, {Morello}, {Mura}, {Narita}, {Nascimbeni}, {Nguyen Tong}, {Noce}, {Oliva}, {Palle}, {Palmer}, {Pancrazzi}, {Papageorgiou}, {Parmentier}, {Perger}, {Petralia}, {Pezzuto}, {Pierrehumbert}, \& {Pillitteri}}]{Tinetti2018}
{Tinetti}, G., {Drossart}, P., {Eccleston}, P., {et~al.} 2018, Experimental Astronomy, 46, 135

\bibitem[{{Trees} \& {Stam}(2019)}]{trees2019}
{Trees}, V.~J.~H. \& {Stam}, D.~M. 2019, \aap, 626, A129

\bibitem[{{Trees} \& {Stam}(2022)}]{trees2022}
{Trees}, V.~J.~H. \& {Stam}, D.~M. 2022, \aap, 664, A172

\bibitem[{{Turbet} {et~al.}(2016){Turbet}, {Leconte}, {Selsis}, {Bolmont}, {Forget}, {Ribas}, {Raymond}, \& {Anglada-Escud{\'e}}}]{turbet2016}
{Turbet}, M., {Leconte}, J., {Selsis}, F., {et~al.} 2016, \aap, 596, A112

\bibitem[{{Vaughan} {et~al.}(2024){Vaughan}, {Birkby}, {Thatte}, {Carlotti}, {Houll{\'e}}, {Pereira-Santaella}, {Clarke}, {Vigan}, {Lin}, \& {Kaltenegger}}]{vaughan2024}
{Vaughan}, S.~R., {Birkby}, J.~L., {Thatte}, N., {et~al.} 2024, \mnras, 528, 3509

\bibitem[{{Vaughan} {et~al.}(2023){Vaughan}, {Gebhard}, {Bott}, {Casewell}, {Cowan}, {Doelman}, {Kenworthy}, {Mazoyer}, {Millar-Blanchaer}, {Trees}, {Stam}, {Absil}, {Altinier}, {Baudoz}, {Belikov}, {Bidot}, {Birkby}, {Bonse}, {Brandl}, {Carlotti}, {Choquet}, {van Dam}, {Desai}, {Fogarty}, {Fowler}, {van Gorkom}, {Gutierrez}, {Guyon}, {Haffert}, {Herscovici-Schiller}, {Hours}, {Juanola-Parramon}, {Kleisioti}, {K{\"o}nig}, {van Kooten}, {Krasteva}, {Laginja}, {Landman}, {Leboulleux}, {Mouillet}, {N'Diaye}, {Por}, {Pueyo}, \& {Snik}}]{vaughan2023}
{Vaughan}, S.~R., {Gebhard}, T.~D., {Bott}, K., {et~al.} 2023, \mnras, 524, 5477

\bibitem[{{Wagner} {et~al.}(2021){Wagner}, {Boehle}, {Pathak}, {Kasper}, {Arsenault}, {Jakob}, {K{\"a}ufl}, {Leveratto}, {Maire}, {Pantin}, {Siebenmorgen}, {Zins}, {Absil}, {Ageorges}, {Apai}, {Carlotti}, {Choquet}, {Delacroix}, {Dohlen}, {Duhoux}, {Forsberg}, {Fuenteseca}, {Gutruf}, {Guyon}, {Huby}, {Kampf}, {Karlsson}, {Kervella}, {Kirchbauer}, {Klupar}, {Kolb}, {Mawet}, {N'Diaye}, {Orban de Xivry}, {Quanz}, {Reutlinger}, {Ruane}, {Riquelme}, {Soenke}, {Sterzik}, {Vigan}, \& {de Zeeuw}}]{Wagner2021}
{Wagner}, K., {Boehle}, A., {Pathak}, P., {et~al.} 2021, Nature Communications, 12, 922

\bibitem[{{Wang} {et~al.}(2022){Wang}, {Fujii}, \& {He}}]{Wang2022}
{Wang}, F., {Fujii}, Y., \& {He}, J. 2022, \apj, 931, 48

\bibitem[{{Way} {et~al.}(2017){Way}, {Aleinov}, {Amundsen}, {Chandler}, {Clune}, {Del Genio}, {Fujii}, {Kelley}, {Kiang}, {Sohl}, \& {Tsigaridis}}]{Way2017}
{Way}, M.~J., {Aleinov}, I., {Amundsen}, D.~S., {et~al.} 2017, \apjs, 231, 12

\bibitem[{{Zieba} {et~al.}(2023){Zieba}, {Kreidberg}, {Ducrot}, {Gillon}, {Morley}, {Schaefer}, {Tamburo}, {Koll}, {Lyu}, {Acu{\~n}a}, {Agol}, {Iyer}, {Hu}, {Lincowski}, {Meadows}, {Selsis}, {Bolmont}, {Mandell}, \& {Suissa}}]{zieba2023}
{Zieba}, S., {Kreidberg}, L., {Ducrot}, E., {et~al.} 2023, \nat, 620, 746

\end{thebibliography}

\begin{appendix}

\section{Cloud averaging over long integration times}
\label{sec:appendix_A}

\begin{figure*}[h]
    \centering
    \includegraphics[width=1\linewidth]{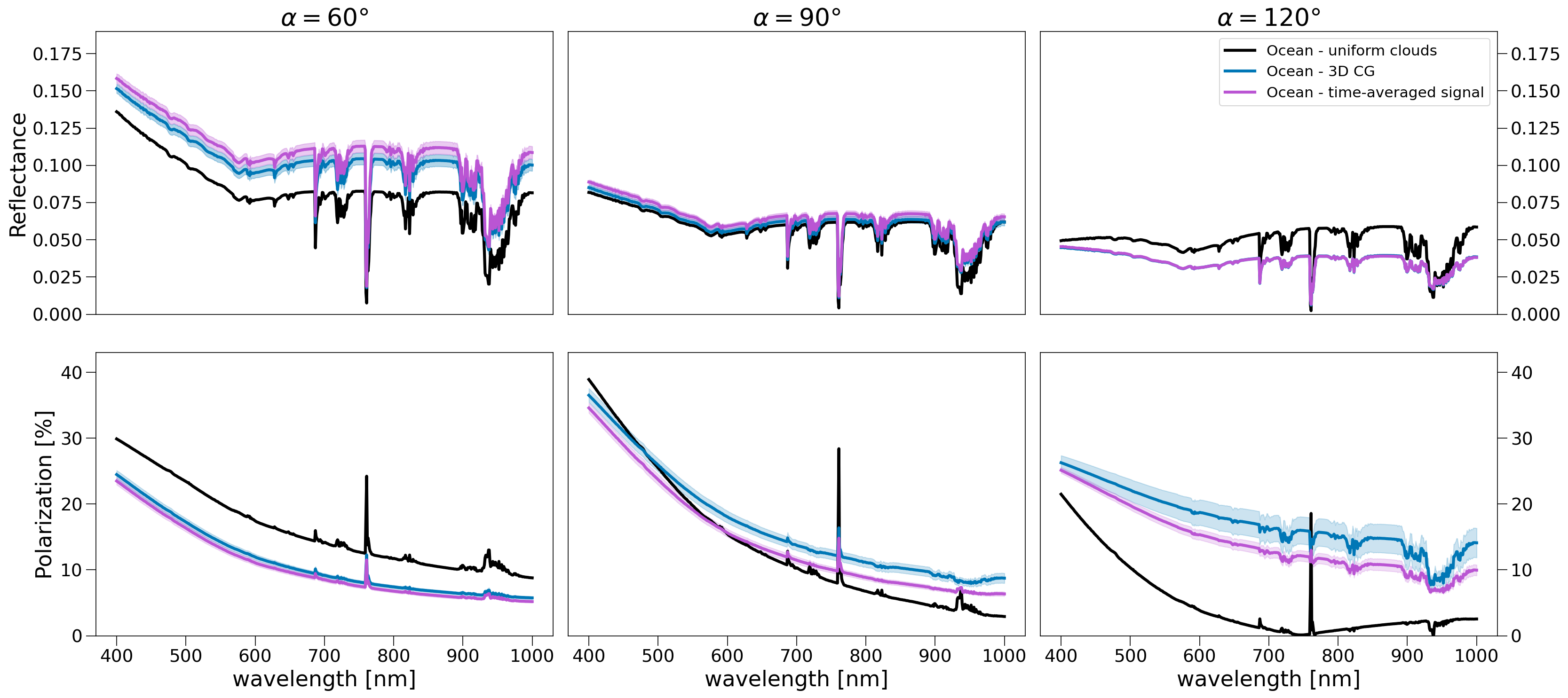}
    \caption{Comparison among spectra in reflected (first row) and polarized light (second row) for an Ocean planet with a uniform cloud layer (black curve), the 3D CG model with 1$\sigma$ variability over 12 months (blue curve), and the time-averaged signal model simulating 8 hours of integration time (purple curve). Different columns refer to different phase angles ($\alpha$): 60, 90, 120\degr.}
    \label{fig:spectra_smeared}
\end{figure*}

\begin{figure*}[h]
    \centering
    \includegraphics[width=1\linewidth]{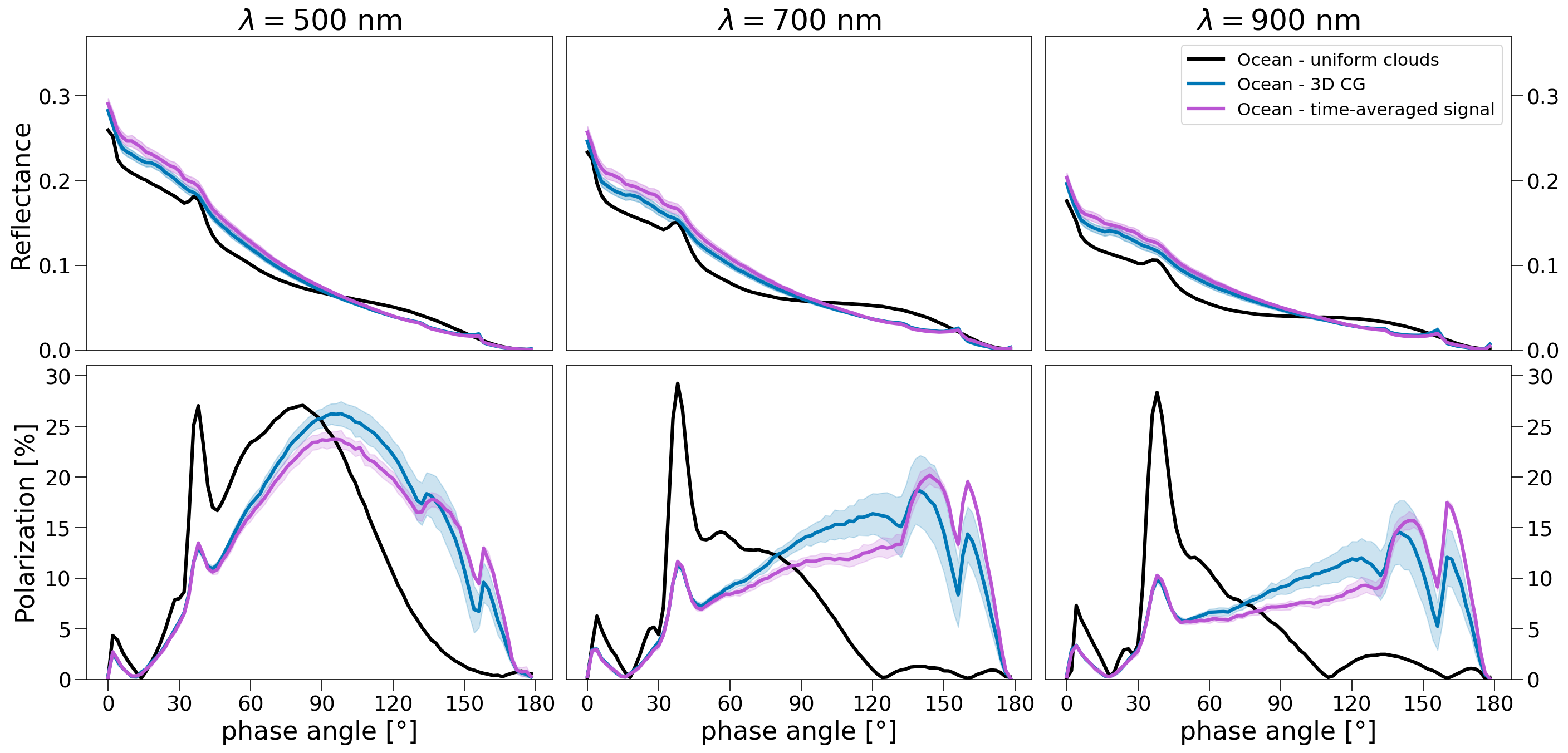}
    \caption{Reflected light (first row) and polarized light (second row) phase curves showing an Ocean planet with a uniform cloud layer (black curve), the 3D CG model with 1$\sigma$ variability over 12 months (blue curve), and the time-averaged signal model simulating 8 hours of integration time (purple curve). Different columns refer to different wavelengths ($\lambda$): 500, 700, 900~nm.}
    \label{fig:phase_smeared}
\end{figure*}

To address the effect of time-averaging the planetary signal over long observational periods, we analyze how evolving cloud patterns influence the resulting spectra and phase curves. This consideration is particularly relevant for future instruments such as ANDES and PCS at the ELT, which will require extended integration times to characterize the golden sample exoplanets. Specifically, we selected one random date from the 12 ERA5 cloud fields used in \cite{Roccetti2025a} to construct the 3D CG 1$\sigma$ cloud spread model. For that date (2023.10.07) we extracted cloud fields from ERA5 at hourly intervals between 16:00 and 23:00 UT, which corresponds to the highest temporal resolution available in ERA5. Using these eight consecutive hourly cloud fields as inputs to the 3D CG model, we ran eight independent simulations. We then computed the average and 1$\sigma$ spread of the resulting outputs, referring to this configuration as the time-averaged signal case. This model is compared against the standard 3D CG 1$\sigma$ spread over 12 months and a uniform cloud model, as described in Sec. \ref{sec:models_definition}, but for an Ocean surface. The impacts on the reflected spectra and phase curves are shown in Figs. \ref{fig:spectra_smeared} and \ref{fig:phase_smeared}, respectively.\\
\noindent We find that the time-averaged signal model (purple curve) closely resembles the 3D CG model (blue curve) across all tested spectra and phase curves. Minor discrepancies with the 3D CG arise because the time-averaged signal model incorporates cloud fields not only over the ocean, but also over different geographic regions during the 8-hour window, starting with the American continent and later extending over the Pacific Ocean. These regional differences explain the observed deviations. Nevertheless, the uniform cloud model (black curve) fails to reproduce the results of the time-averaged signal model, showing significant discrepancies, particularly at large phase angles in reflectance (Fig. \ref{fig:spectra_smeared}, top-left panel), and in polarization, both in the spectral slope at $\alpha$ = 90\degr (bottom-center panel) and especially at large phase angles (bottom-right panel).\\
\noindent In the comparison of phase curves shown in Fig. \ref{fig:phase_smeared}, we again observe some small deviations between the time-averaged signal model and the 3D CG model, as well as with the uniform cloud model. In reflectance, these differences are not strongly pronounced; however, we note an overestimation of the cloudbow feature with uniform clouds and an underestimation of the continuum level. In polarization (second row), the differences become much more significant. The cloudbow feature is substantially overestimated in the uniform cloud model. At wavelengths of 700 and 900~nm, the uniform model fails to reproduce the polarization continuum at phase angles greater than 90\degr, due to its inability to capture the specular reflection from the ocean glint, an effect that strongly polarizes light. In contrast, the time-averaged signal model still exhibits the ocean glint signature in the polarization continuum. This is because the patchy nature of the evolving cloud fields allows photons to reach the surface at all times, even under changing weather conditions. \\
\noindent Thus, even when averaging cloud coverage over long integration times, it is important to account for realistic cloud patchiness, as it has a significant impact on both reflected and polarized light signals, effects that cannot be replicated by homogeneous cloud models.

\end{appendix}

\end{document}